\documentclass[article,twocolumn,floatfix]{revtex4}

\usepackage{graphicx} 
\usepackage{subfig}
\usepackage{color}

\newcommand{\ket}[1]{\left| #1 \right\rangle}

\renewcommand{\Re}{\mathrm{Re}}
\newcommand{\be}{\begin{equation}}
\newcommand{\ee}{\end{equation}}
\newcommand{\bea}{\begin{eqnarray}}
\newcommand{\eea}{\end{eqnarray}}

\usepackage{dcolumn}
\usepackage{bm}
\usepackage{amssymb,amsmath}
\usepackage{color}
\usepackage{float}
\usepackage{tikz}
\usetikzlibrary{arrows}

\definecolor{DarkGreen}{rgb}{0,0.6,0.2}

\begin{document}
\title{Controlling tripartite entanglement among optical cavities by reservoir engineering}
\author{Imran M. Mirza }
\affiliation{Oregon Center for Optics and Department of Physics\\University of Oregon\\
Eugene, OR 97403}
\begin{abstract}
We study how to control the dynamics of tripartite entanglement among optical cavities using non-Markovian baths. In particular, we demonstrate how the reservoir engineering through the utilization of non-Markovian baths with different types of Lorentzian and ohmic spectral densities can lead to an entanglement survival for longer times and in some cases considerable regain of seemingly lost entanglement. Both of these behaviors indicate a better sustainability of entanglement (in time) compared to the usual Markovian bath situations which assumes a flat spectrum of the bath around the system resonant frequency. Our scheme shows these effects in the context of optical cavities starting off in a maximally entangled W and Greenberger-Horne-Zeilinger (GHZ) tripartite states. In Lorentzian cases we find that the far detuned double Lorentzian baths with small coupling strengths and for ohmic type baths super-ohmic environments with smaller cutoff frequencies are the best candidates for preserving entanglement among cavities for significant amount of time. A non-Markovian quantum jump approach is employed to understand the entanglement dynamics in these cases, especially to recognize the collapse and revival of the entanglement in both W and GHZ states. 
\end{abstract}
\maketitle

\section{Introduction}
The problem of maintaining entanglement \cite{horodecki2009quantum} among quantum systems that are open to their environments is one of the most challenging issues in quantum information processing today \cite{dodd2004disentanglement}. It is well known that, even if the interactions are weak, the environment (bath/reservoir) dramatically alters the dynamics of the quantum systems, which results in an irreversible loss of information from the system to the environment: a phenomena more commonly known as the decoherence  \cite{zurek2003decoherence}. As a result, this lost information doesn't play any further role in the dynamics of the system and the system loses its pure quantum features (in particular entanglement). \\
The successful generation and longer survival of entanglement is the central ingredient to a variety of quantum information processing tasks such as quantum teleportation \cite{bouwmeester1997experimental}, quantum dense coding \cite{bouwmeester2000physics}, quantum cryptography \cite{jennewein2000quantum} and quantum computing \cite{jozsa1997entanglement}. Entanglement also plays a crucial role in the foundations of quantum mechanics and in the non-classical behaviour of correlated many-body quantum systems \cite{amico2008entanglement}. Based on this broad range of applications, devising novel ways of preserving entanglement among open quantum systems (for considerable times) is of central importance for the  practical applications of multipartite quantum systems in future quantum technologies. In order to control and reduce the effects of decoherence, in recent years reservoir engineering techniques are developed at a practical level, which are now revising the role of the system-environment interactions \cite{myatt2000decoherence, busch2010reservoir, verstraete2009quantum}. These techniques are leading to new ways to manage the properties of environment which can help in sustaining the pure quantum features of the system for extended times.\\ 

Motivated by these techniques, in this work we address this decisive question that once the entanglement among open quantum systems is established, how we can control/save it for a sufficient amount of time so that the quantum protocol (related to the quantum communication or quantum computation under consideration) can be successfully performed. We answer this question in the context of tripartite entanglement which is assumed to be generated among three open and independent optical cavities. Our choice of studying tripartite entanglement is based on the fact that, despite of intense research in entanglement for last two decades, at present our full understanding regarding entanglement is merely limited to bipartite systems. Although there are many criteria introduced to quantify multipartite entanglement of mixed and pure states \cite{horodecki2009quantum}, a full framework is still lacking. One particular class of multipartite entangled states which is relatively easy to study is the tripartite entangled states \cite{roos2004control}. W and GHZ states are the celebrated examples belonging to this class of entangled states. Along with many applications in quantum information processing (as indicated in the references above) the tripartite entangled states may also provide a platform to understand the differences and similarities in the behavior of multipartite and bipartite entanglement at the fundamental level. \\

The basic idea that we are going to exert in the present study will be to employ the technique of reservoir engineering through the utilization of non-Markovian baths (with different spectral profiles) for freezing the entanglement among tripartite-cavity system. More precisely, we'll investigate the type of non-Markovian baths that are most suitable for the entanglement preservation from environmental decoherence. In recent years non-Markovian baths have attracted considerable attention due to their diverse applications in the foundations of quantum theory, in quantum computation and quantum communications \cite{zhang2012general, lorenzo2013geometrical}. The reason for utilizing non-Markovian baths for entanglement survival purposes, lies in the fact that in the standard approach of studying open quantum systems, environments are usually assumed to be memory-less (Markovian) on the time scales of the system dynamics. As a result, environment always act as a sink for the information contained in the system. It is well known that the bipartite entangled systems suffer entanglement sudden death (ESD) \cite{yu2009sudden,salles2008experimental} due to this effect.\\

 Such a fast decay of entanglement is not desirable for several quantum information and quantum computation protocols where entanglement plays a key role \cite{kim2012protecting}.  
 Hence we employ non-Markovian environments where we'll have more freedom of choosing environments with different spectral functions as well as we can identify different ways of probing the environment and environment-system interaction such that entanglement can be sustained for longer times. In here, we are going to consider the non-Markovian environmental models which have experimental relevance too, such as in  condensed matter \cite{dattagupta2004dissipative}, biological \cite{devault1980quantum} and quantum optical \cite{madsen2011observation, hoeppe2012direct} systems (to name a few). Except one case of sub-ohmic spectral function, we find that using non-Markovian baths with different Lorentzian and ohmic type spectra can better perform in the longer entanglement survival compared to usual Markovian baths. In particular, weakly interacting far detunned double Lorentzian baths and super-ohmic baths with smaller cut-off frequencies establish an entanglement which is more robust against environmental losses.  Moreover, a non-Markovian quantum jump approach (NMQJA as introduced by Piilo et.al \cite{piilo2008non}) is found to be sufficient to explain the loss and gain of entanglement in terms of positive and negative cycles of time-dependent decay rates. Here we would to mention that the bipartite and tripartite qubit systems dynamics when coupled to Markovian and non-Markovian baths is already a studied subject \cite{dukalski2012tripartite, man2010entanglement, bina2010tripartite}. The main novelty of our work is the consideration of various environmental models (with finite temperratures) in this context and the application of NMQJA to understand the entanglement dynamics, which to our knowledge has not been reported previously for the system under consideration.\\

The paper is organized as follows: in Sec. II, we model the system by writing down the Hamiltonian and include the dissipative dynamics of the system by coupling the cavities to their respective baths with different spectral densities. In Sec. III, we present results and discussion where we'll first consider the entanglement dynamics when the optical cavities (initially either in W or in GHZ state) are coupled to a Markovian bath. Effect of photon hopping between cavities will also be considered. This section will be useful for later discussions on a comparison between entanglement dynamics for Markovian and non-Markovian baths cases. Next we study the effect of non-Markovian baths on the cavity based tripartite entanglement dynamics of W and GHZ states. Finally in Sec. V we'll summarize the results and report conclusions of this work.

\section{Hamiltonian and Dissipative Dynamics}
System under study comprises of three independent and open optical cavities. For simplicity it is assumed that each cavity has a single isolated resonant mode, which are given by frequencies $\omega_{c1},\omega_{c2}$ and $\omega_{c3}$ in the first, second and third cavity respectively. Destruction of photon in these cavities is described by the annihilation operator $\hat{a}_{i}$ $\forall i = 1,2,3$ (throughout the article we'll use same range of values for index $i$). Under rotating wave approximation, the total/global Hamiltonian describing the system Hamiltonian ($\hat{H}_{S}$), bath Hamiltonian ($\hat{H}_{B}$) and system-bath interaction Hamiltonian ($\hat{H}_{SB}$) is given by:
\begin{equation}\label{H}
\begin{split}
&\hat{H}_{sys} = \hat{H}_{S}+\hat{H}_{B}+\hat{H}_{SB}=\sum_{i=1}^{3}\Bigg(\hbar\omega_{ci}+\sum_{k_{i}}\hbar\omega_{k_{i}}\hat{b}^{\dagger}_{k_{i}}\hat{b}_{k_{i}}\\
&+\sum_{k_{i}}\hbar g_{k_{i}}\lbrace \hat{a}_{i}\hat{b}^{\dagger}_{k_{i}}+\hat{a}^{\dagger}_{i}\hat{b}_{k_{i}}\rbrace\Bigg)
\end{split}
\end{equation}
Baths are assumed to have a continuum of modes while $\hat{b}_{k_{i}}$ operator annihilates a photon in the $k$th mode (frequency $\omega_{k_{i}}$) of the $i$th bath. The system (cavity) mode interaction with the bath modes is characterized by the real coupling rate $g_{k_{i}}$ without loss of generality. Non-vanishing commutation relations are given by: $[\hat{a}_{i},\hat{a}^{\dagger}_{j}] = \delta_{ij} $,  $[\hat{b}_{k_{i}},\hat{b}^{\dagger}_{k_{j}}] = \delta_{ij}, \forall i = 1,2,3$ and $j = 1,2,3$. \\
When the cavities are coupled with non-Markovian baths, the system dynamics can be described in terms of the so-called Exact Master Equations (EME)\cite{vacchini2010exact}. Basic assumption in these types of Master equations is the dependence of the time derivative of the system density operator on the state of the system at the present time. Although the general structure of EME resembles the Lindblad Master equations but due to the presence of time-dependent functions (which are related to the photon decay rates and energy shifts \cite{mazzola2010phenomenological}) the EME are generally non-Lindblad per se. It turns out that these time-dependent decay rates are responsible for the production of memory effects (non-Markovianity) in EME (we'll notice this fact in the context of entanglement evolution in later sections).\\

In interaction picture, EME for present systems takes the following form \cite{breuer2002theory,ali2010decoherence}:
\begin{equation}\label{EME}
\begin{split}
& \frac{d\rho_{s}(t)}{dt} =\sum_{i=1}^{3}\Bigg[\alpha_{i}(t)\Bigg(\hat{a}^{\dagger}_{i}\hat{\rho}_{s}(t)\hat{a}_{i}-\hat{\rho}_{s}(t)\hat{a}_{i}\hat{a}^{\dagger}_{i}\Bigg)\\
&+\alpha^{\ast}_{i}(t)\Bigg(\hat{a}^{\dagger}_{i}\hat{\rho}_{s}(t)\hat{a}_{i}-\hat{a}_{i}\hat{a}^{\dagger}_{i}\hat{\rho}_{s}(t)\Bigg)+\beta_{i}(t)\Bigg(\hat{a}_{i}\hat{\rho}_{s}(t)\hat{a}^{\dagger}_{i}-\\
&\hat{a}^{\dagger}_{i}\hat{a}_{i}\hat{\rho}_{s}(t)\Bigg)+\beta^{\ast}_{i}(t)\Bigg(\hat{a}_{i}\hat{\rho}_{s}(t)\hat{a}^{\dagger}_{i}-\hat{\rho}_{s}(t)\hat{a}^{\dagger}_{i}\hat{a}_{i}\Bigg)\Bigg]
\end{split}
\end{equation}

The time-dependent functions $\alpha_{i}(t)$ and $\beta_{i}(t)$ are related to the non-Markovian baths' spectral densities $J_{i}(\omega)$ in the following way:
\begin{subequations}
\begin{eqnarray}
\hspace{-5mm}\alpha_{i}(t)=\int_{0}^{t}dt_{1}\int_{0}^{\infty}d\omega J_{i}(\omega)\overline{N}(\omega)e^{i(\omega_{ci}-\omega)(t-t_{1})}\\
\label{alphat}
\hspace{-5mm}\beta_{i}(t)=\int_{0}^{t}dt_{1}\int_{0}^{\infty}d\omega J_{i}(\omega)(\overline{N}(\omega)+1)e^{i(\omega_{ci}-\omega)(t-t_{1})}\label{betat}
\end{eqnarray}
\end{subequations}
while bath spectral densities $J_{i}(\omega)$ are in turn dependent on the cavity-reservoir interaction rate $g_{k_{i}}$ through $J_{i}(\omega)=\int dt \sum_{k}g^{2}_{k_{i}}e^{i(\omega-\omega_{k})t}$ \cite{gardiner2004quantum}. For the sake of simplicity all baths are assumed to have same average number of photons $\overline{N}(\omega)$ as given by the Planck's distribution $\frac{1}{e^{\hbar\omega/k_{B}T}-1}$ while $k_{B}$ is the Boltzmann constant and $T$ is the absolute temperature of the bath.\\
 
 Choosing the basis set according to the truncated Hilbert Space ($\mathcal{H}$) relevant to the present three cavity system: $\lbrace \ket{000}\rightarrow\ket{1}, \ket{100}\rightarrow\ket{2}, \ket{010}\rightarrow\ket{3}, \ket{001}\rightarrow\ket{4}, \ket{110}\rightarrow\ket{5}, \ket{101}\rightarrow\ket{6}, \ket{011}\rightarrow\ket{7}, \ket{111}\rightarrow\ket{8}\rbrace$ (while the first, second and third slots in the ket represent the number of photons in the first, second and third cavities respectively), and assuming all the time-dependent functions to be the same i.e. $\alpha_{1}(t)=\alpha_{2}(t)=\alpha_{3}(t)=\alpha(t)$ and $\beta_{1}(t)=\beta_{2}(t)=\beta_{3}(t)=\beta(t)$, one can calculate the time evolution of density matrix elements. For W state, initially non-vanishing density matrix elements are: $\rho_{22}(t=0)= \rho_{23}(t=0)=\rho_{24)}(t=0)=\rho_{32}(t=0)=\rho_{33}(t=0)=\rho_{34}(t=0)=\rho_{42}(t=0)=\rho_{43}(t=0)=\rho_{44}(t=0)=1/3$.
For GHZ state initial conditions are: $\rho_{11}(t=0)=\rho_{18}(t=0)=\rho_{81}(t=0)=\rho_{88}(t=0)=1/2 $. Full solution of the density operator $\rho_{s}(t)$ is expressed in terms of following matrix:\\
\begin{equation}
\rho_{s}(t)=
\begin{pmatrix}
\rho_{11}& \rho_{12}& \rho_{13}& \rho_{14}& \rho_{15}& \rho_{16}& \rho_{17}& \rho_{18}&\\
\rho_{21}& \rho_{22}& \rho_{23}& \rho_{24}& \rho_{25}& \rho_{26}& \rho_{27}& \rho_{28}&\\
\rho_{31}& \rho_{32}& \rho_{33}& \rho_{34}& \rho_{35}& \rho_{36}& \rho_{37}& \rho_{38}&\\
\rho_{41}& \rho_{42}& \rho_{43}& \rho_{44}& \rho_{45}& \rho_{46}& \rho_{47}& \rho_{48}&\\
\rho_{51}& \rho_{52}& \rho_{53}& \rho_{54}& \rho_{55}& \rho_{56}& \rho_{57}& \rho_{58}&\\
\rho_{61}& \rho_{62}& \rho_{63}& \rho_{64}& \rho_{65}& \rho_{66}& \rho_{67}& \rho_{68}&\\
\rho_{71}& \rho_{72}& \rho_{73}& \rho_{74}& \rho_{75}& \rho_{76}& \rho_{77}& \rho_{78}&\\
\rho_{81}& \rho_{82}& \rho_{183}& \rho_{84}& \rho_{85}& \rho_{86}& \rho_{87}& \rho_{88}&
\end{pmatrix}
\end{equation}

In general, non-Markovian baths can have any type of spectral profile. In present study, we'll consider different (and experimental relevant) types of environmental models having: single Lorentzian ($J_{{\rm SL}}$), double Lorentzian ($J_{{\rm DL}}$), band-gap Lorentzian ($J_{{\rm BL}}$), sub-ohmic ($J_{{\rm SBO}}$), ohmic ($J_{{\rm O}}$) and super-ohmic ($J_{{\rm SPO}}$) spectral densities as expressed below and shown in Fig.~1: 
 \begin{subequations}
 \begin{align}
J_{{\rm SL}}(\omega) = \frac{\alpha_{{\rm L}}}{2\pi}\frac{\Gamma^{2}}{(\omega-\omega_{bc})^{2}+(\Gamma/2)^{2}},
\end{align}\\
\vspace{-8mm}
\begin{align}
J_{{\rm DL}}(\omega) = W_{D1}\Bigg(\frac{\alpha_{{\rm L1}}}{2\pi}\frac{\Gamma^{2}_{1}}{(\omega-\omega_{bc})^{2}+(\Gamma_{1}/2)^{2}}\Bigg)\nonumber\\
+W_{D2}\Bigg(\frac{\alpha_{{\rm L2}}}{2\pi}\frac{\Gamma_{2}^{2}}{(\omega-\omega_{bc})^{2}+(\Gamma_{2}/2)^{2}}\Bigg),
\end{align}\\
\vspace{-8mm}
\begin{align}
J_{{\rm BL}}(\omega) = W_{B1}\Bigg(\frac{\alpha_{{\rm L1}}}{2\pi}\frac{\Gamma^{2}_{1}}{(\omega-\omega_{bc})^{2}+(\Gamma_{1}/2)^{2}}\Bigg)\nonumber\\
-W_{B2}\Bigg(\frac{\alpha_{{\rm L2}}}{2\pi}\frac{\Gamma^{2}_{2}}{(\omega-\omega_{bc})^{2}+(\Gamma_{2}/2)^{2}}\Bigg),
\end{align}\\
\vspace{-8mm}
\begin{align}
J(\omega) = \alpha\omega_{cut}^{1-s}\omega^{s} e^{-\omega/\omega_{cut}}.
\end{align}
 \end{subequations}
 
\begin{figure}
\begin{center}
\includegraphics[width=3in, height=2in]{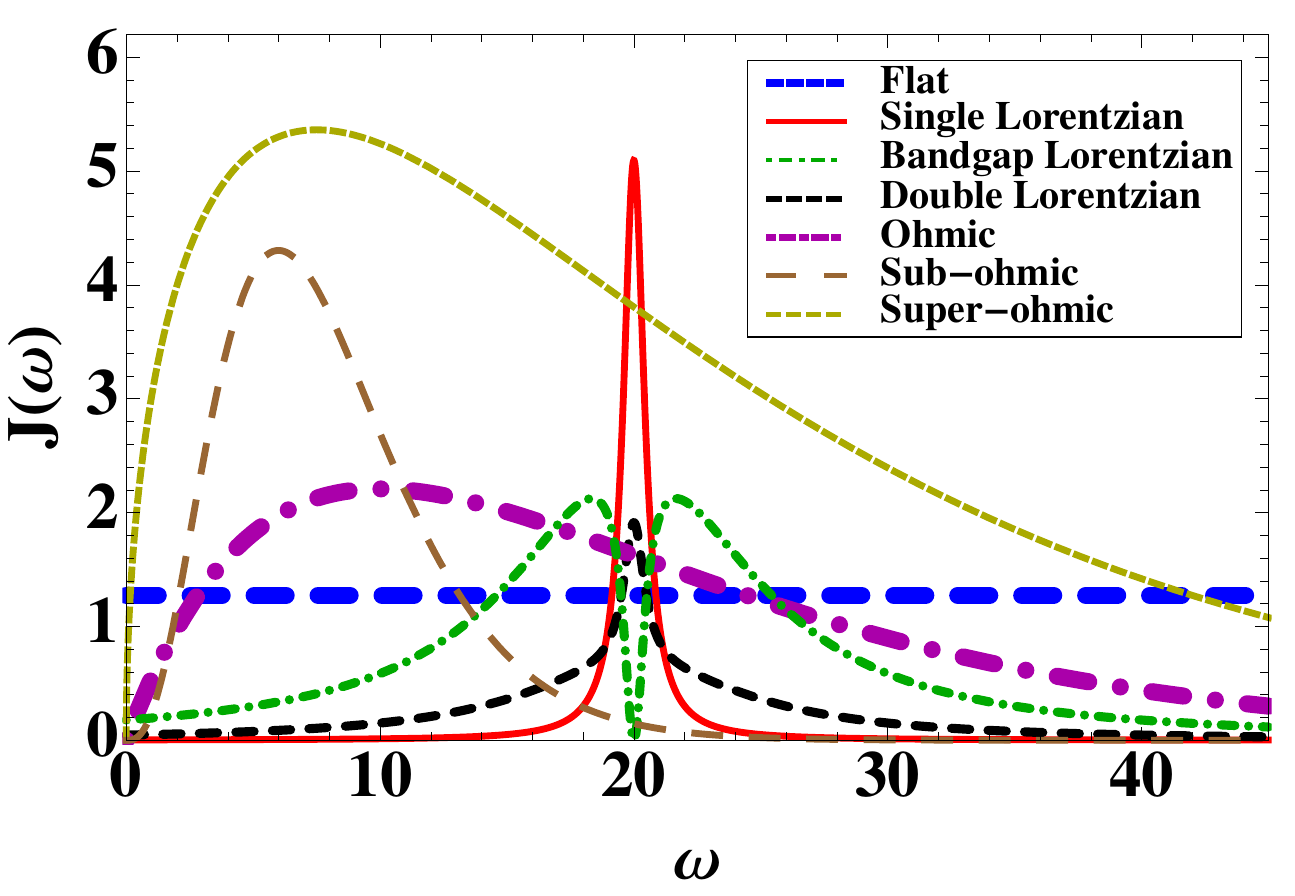}
\captionsetup{
  format=plain,
  margin=1em,
  justification=raggedright,
  singlelinecheck=false
}
  \caption{Spectral density functions describing different environmental models. Parametere used are: For Loretzian functions (single, double and band-gap Lorentzian) $\alpha_{L} = \alpha_{L1}=\alpha_{L2} = 2$, $\Gamma =\Gamma_{1} =\Gamma_{2} = 1$, $\omega_{bc} = 20$, $W_{D1} = W_{D2} = 1/2$ and $W_{B1} = 2, W_{B2} = 1$. For ohmic type densities we have used $\omega_{cut} = 15, 10, 2$ and $\alpha = 1, 0.6, 0.1$ for sub-ohmic, ohmic and super-ohmic functions respectively. Note that we have also plotted the flat spectrum density which is commonly assumed in the case of Markovian baths.}\label{Fig2}
\end{center}
\end{figure}
The non-Markovian bath with a Lorentzian spectral density can be experimentally realized by coupling the quantum system with a high finesse (leaky) optical cavity (which we'll call the bath cavity). In pseudomode picture of non-Markovian dynamics \cite{mazzola2009pseudomodes} system's cavity mode couples with the pseudomode of the bath cavity which in turn couples with the outside Markovian environment. In single Lorentzian density, $\alpha_{L}$, $\Gamma$ and $\omega_{bc}$ express the coupling between the bath cavity and the outside environment, width of the spectral distribution and the resonant frequency of bath cavity respectively. In double (band-gap) Lorentzian we have summed (differed) two Lorentzians with the parameters (describing each Lorentzian function) having the same meaning as in the case of single Lorentzian model. In addition to these parameters, in double Lorentzian model both Lorentzians are weighted by the condition: $W_{D1}+W_{D2}=1$ and in band-gap Lorentzian model the conditions $W_{B1}-W_{B2}=1$ and $\Gamma_{2}<\Gamma_{1}$ are respected to ensure positivity of the spectral density as a function of $\omega$.
To examine a bath which considers an even broader range of frequencies, in last example (Eq.5(d)) we have summarized sub-ohmic, ohmic and super-ohmic spectral densities, by introducing spectral parameter $s$. Note that in this equation by choosing $s=1/2, 1$ and $s=3$, we can obtain $J(\omega)\longrightarrow J_{{\rm SBO}}(\omega),J(\omega)\longrightarrow J_{{\rm O}}(\omega)$ and $J(\omega)\longrightarrow J_{{\rm SPO}}(\omega$) respectively. To cut off the higher frequencies (to avoid divergences) $\omega_{cut}$  is introduced in the spectral density $J(\omega)$ while $\alpha$ represents a dimensionless coupling constant.

\section{Results and Discussion}
\subsection{Entanglement dynamics under Markovian baths}
\subsubsection{Without photon hoping}
Different experimental techniques and theoretical proposals are developed for the successful generation of tripartite entanglement among quantum systems. Four wave mixing in cold atomic gases \cite{wen2010tripartite}, spatial mode parametric down conversion \cite{daems2010tripartite}, anisotropic exchange interactions \cite{galiautdinov2008maximally}, single spins in diamond \cite{neumann2008multipartite} and scheme based on cavity quantum electrodynamics (CQED) (photons passing through optical cavities) \cite{guo2002scheme1}, are few prominent examples.\\
To highlight the differences between Markovian and non-Markovian entanglement dynamics, in this section we address the Markovian bath case when three cavity system is initially prepared in a tripartite entangled state (either W or GHZ state). For simplicity we'll begin by assuming no direct or indirect coupling among cavities and among baths as well. With a single photon restriction in each cavity and following the standard reservoir theory \cite{breuer2002theory}, the dynamics of such a two-level three-cavity system can be described by the well known Lindblad Master equation:
\begin{equation}
\begin{split}
&\frac{d\rho_{s}(t)}{dt} =\sum_{i=1}^{3}\Bigg(\kappa_{i}(\overline{N}+1)(\hat{a}_{i}\hat{\rho}_{s}(t)\hat{a}^{\dagger}_{i}-\frac{1}{2}\hat{a}^{\dagger}_{i}\hat{a}_{i}\hat{\rho}_{s}(t)-\\
&\frac{1}{2}\hat{\rho}_{s}(t)\hat{a}^{\dagger}_{i}\hat{a}_{i})+\kappa_{i}\overline{N}(\hat{a}^{\dagger}_{i}\hat{\rho}_{s}(t)\hat{a}_{i}-\frac{1}{2}\hat{a}_{i}\hat{a}^{\dagger}_{i}\hat{\rho}_{s}(t)-\frac{1}{2}\hat{\rho}_{s}(t)\hat{a}_{i}\hat{a}^{\dagger}_{i})\Bigg)
\end{split}
\end{equation}
Note that one can also obtain the above Master equation directly from Eq.[{\ref{EME}}] by setting $J_{i}(\omega) = \kappa_{i}/2\pi$ and $\overline{N}_{i}(\omega)=\overline{N}_{i}\equiv\frac{1}{e^{\hbar\omega_{ci}/k_{B}T}-1}$. The decay rate of energy from the ith cavity is denoted by $\kappa_{i}$ and $\overline{N}$ identifies the temperature of the heat bath which is assumed to be same for all baths.  The density matrix $\hat{\rho}(t)$ (describing the state of the optical cavities) can be used to quantify the amount and temporal evolution of entanglement. There are several different measures of entanglement that can be used to quantify bipartite and multipartite entanglement \cite{horodecki2009quantum}. One measure which we are going to use here is called the negativity ($\mathcal{N}$), which follows from the Peres-Horodecki Positive Partial Transpose (PPT) criterion \cite{peres1996separability}, and it can be calculated for systems of arbitrary Hilbert space dimensions. The range of $\mathcal{N}$ lies between $0$ and $1$ where $1$ refers to maximum entanglement and $0$ to completely separable states. Negativity is defined as:
\begin{equation}\label{Nega}
\mathcal{N}(t) = \max\Bigg(0,-2\sum_{i}\lambda_{i}\Bigg)
\end{equation}
where the sum is taken over the negative eigenvalues $\lambda_{i}$ of the partially transposed density matrix $\hat{\rho}^{PT}_{s}$. Partial transposition is taken with respect to one of the cavities only and here we'll perform it with respect to the first cavity. We find that (independent of bath temperature ($\overline{N}$)), among eight eigenvalues of $\rho^{PT}_{s}$ always only one eigenvalue turns out to be negative. For an initial preparation of the system in W state:
\begin{equation}
\ket{\Psi(t=0)}_{W}=\frac{1}{\sqrt{3}}(\ket{100}+\ket{010}+\ket{001})
\end{equation}
 we find the following analytic form of Negativity:
\begin{equation}
\mathcal{N}_{W} = {\rm max}\Bigg[0, (e^{-\kappa t}-1)(1+(1+\frac{8}{9} e^{-2\kappa t}(e^{-\kappa t}-1)^{-2}))\Bigg]
\end{equation}
And for a GHZ initial state:
\begin{equation}
\ket{\Psi(t=0)}_{GHZ}=\frac{1}{\sqrt{2}}(\ket{000}+\ket{111})
\end{equation}
Negativity turns out to be:
\begin{equation}
\begin{split}
&\mathcal{N}_{GHZ} = {\rm max}\Bigg[0,(\frac{1}{3} e^{-\kappa t}(e^{-\kappa t}-1)+\frac{1}{3}e^{-3\kappa t}(4e^{3\kappa t}+\\
&\frac{1}{3}(4-12e^{\kappa t}+13e^{2\kappa t}-10e^{3\kappa t}+e^{4\kappa t})))\Bigg]
\end{split}
\end{equation}
\begin{figure*}[t]
\begin{center}
\begin{tabular}{cccc}
\subfloat{\includegraphics[width=5cm,height=4cm]{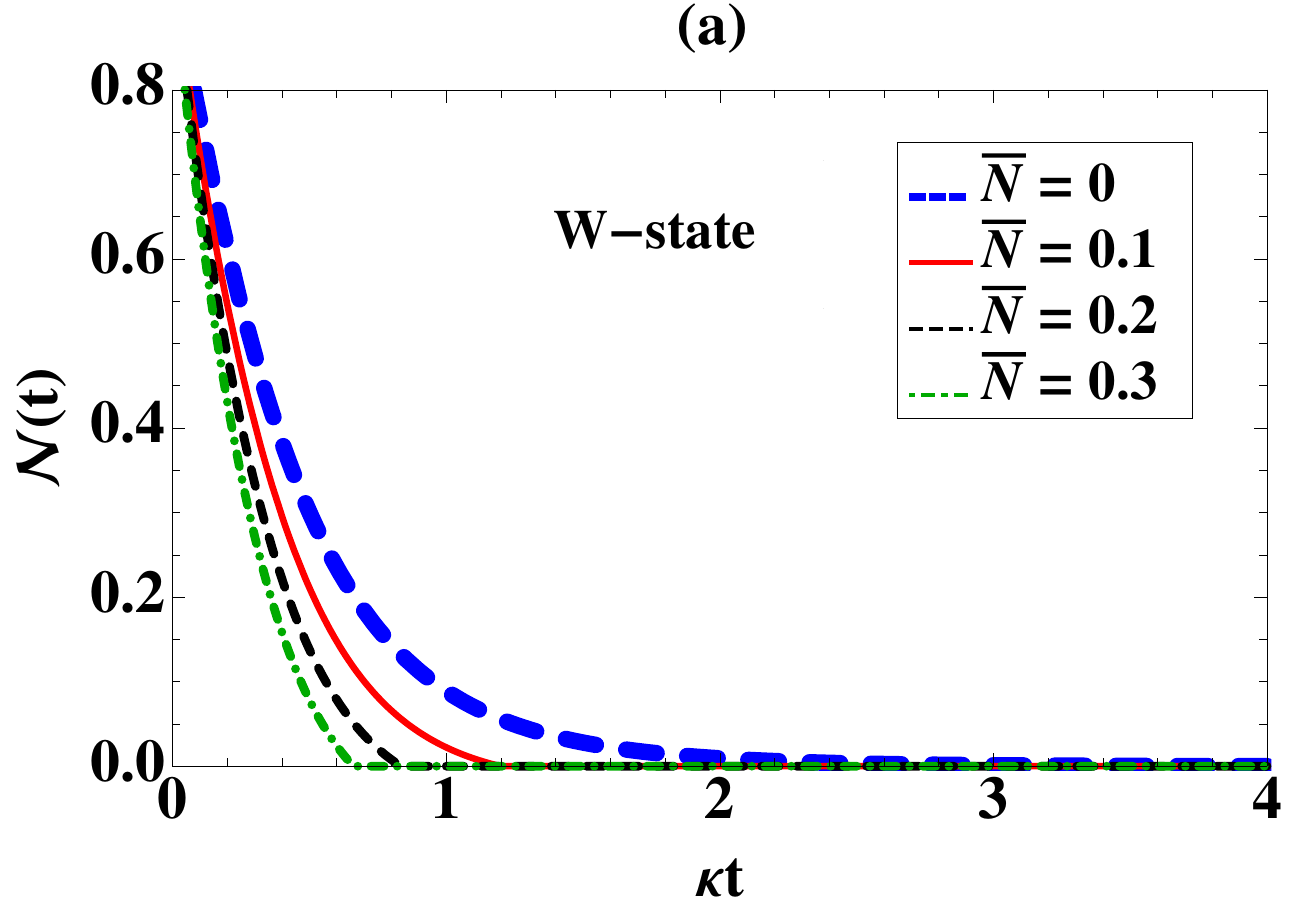}} & 
\subfloat{\includegraphics[width=5cm,height=4cm]{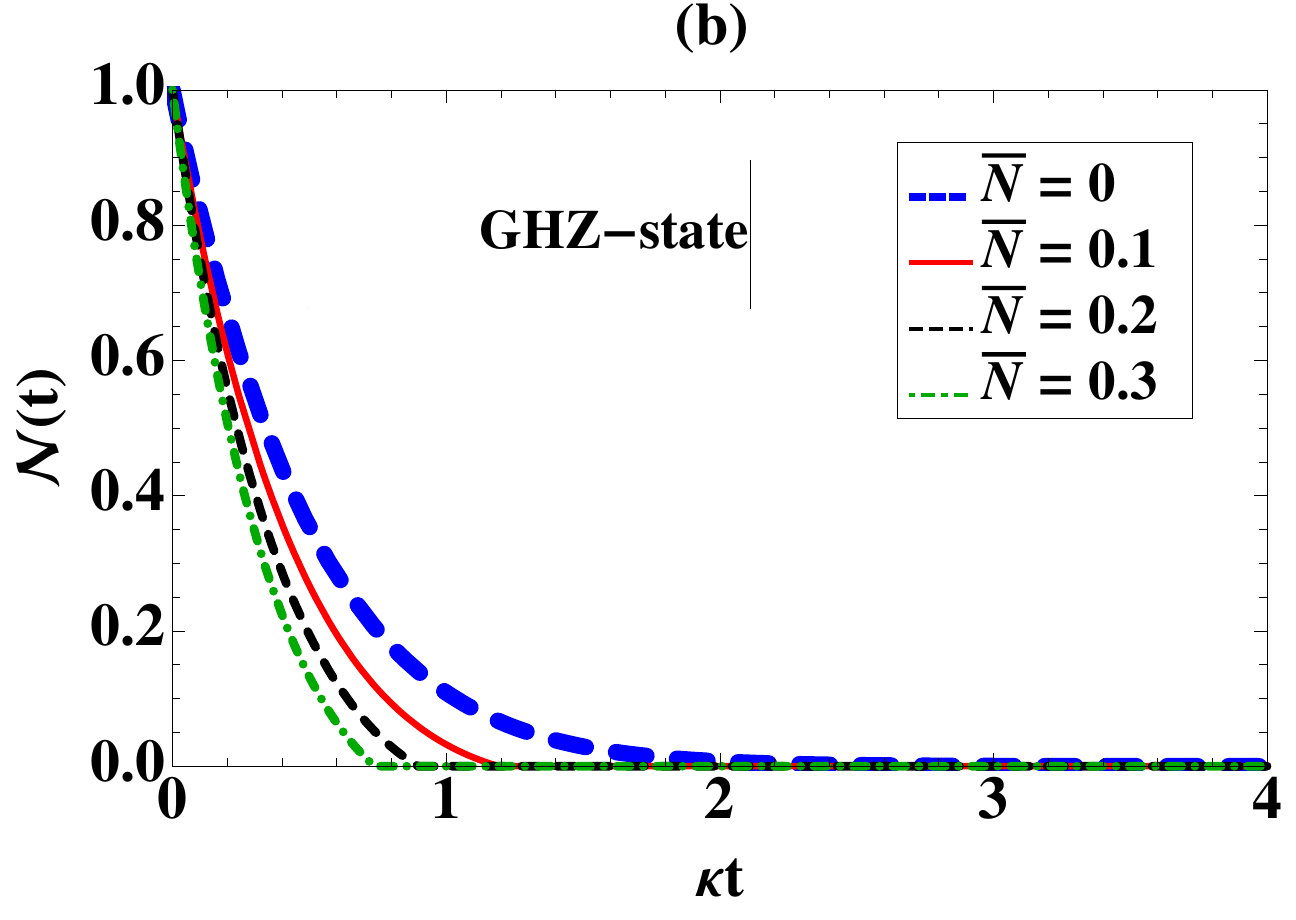}}& 
\subfloat{\includegraphics[width=5cm,height=4cm]{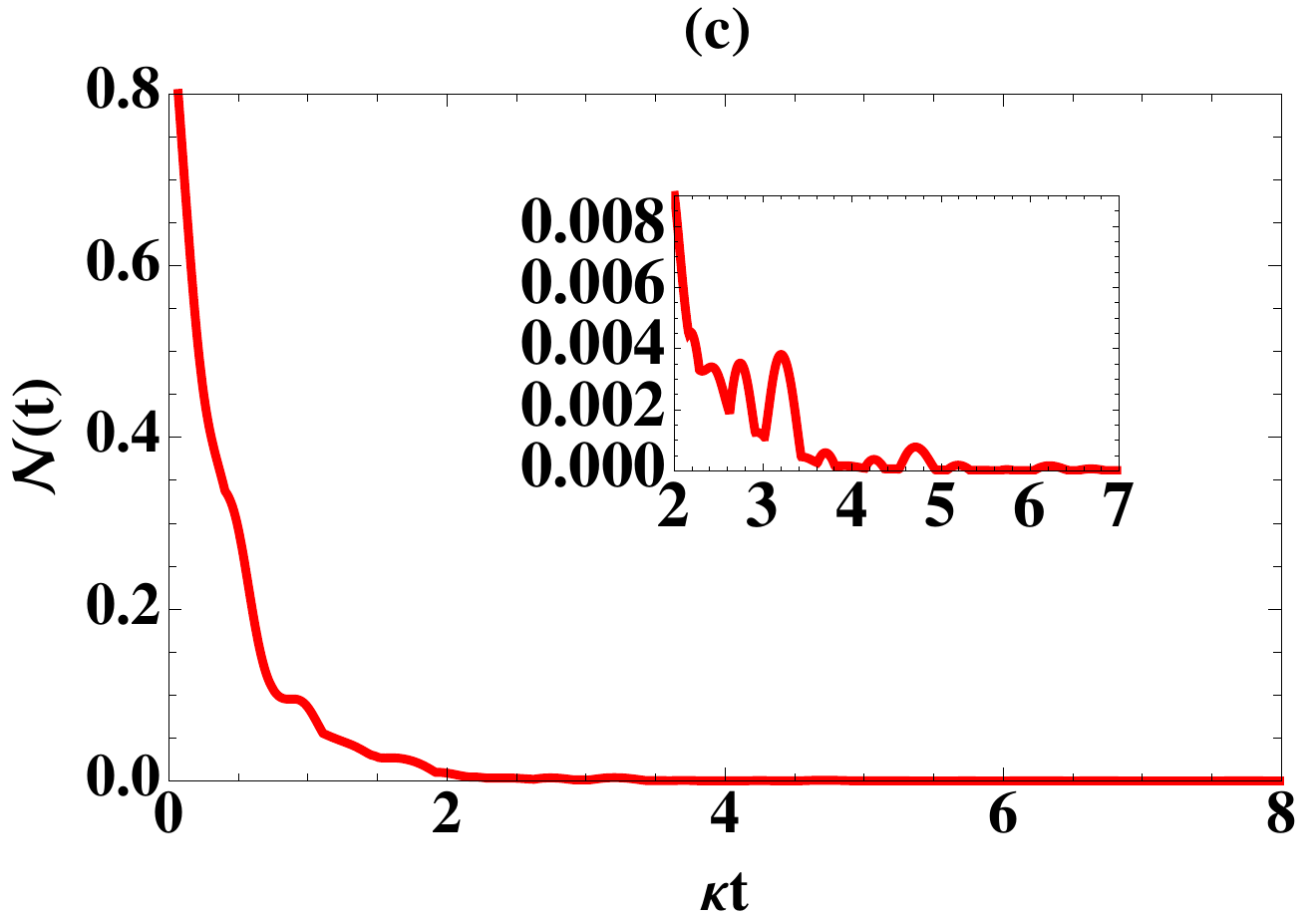}}\\
\end{tabular}
\captionsetup{
  format=plain,
  margin=1em,                          
  justification=raggedright,
  singlelinecheck=false
}
\caption{Time evolution of tripartite entanglemnet among optical cavities characterized by negativity, when cavities were initially prepared in a maximally entangled (a) W and (b) GHZ state excluding the effect of photon hoping. (c) W-state entanglement evolution at zero temperature when a two-way strong photon hoping among cavities is included ($\xi_{12}=\xi_{23}=5\kappa$ with $\kappa_{1}=\kappa_{2}=\kappa_{3}\equiv\kappa$). Note that we have shown the result only for the case of W-state, but the main points of this plot appear to be the identical for an initial GHZ state as well.} \label{Fig2}
\end{center}
\end{figure*}
In FIG.~2(a),-(b) we have plotted Negativity for both W and GHZ states with varying values of $\overline{N}$. In both cases we notice an asymptotic and irretrievable loss of entanglement due to the cavities interaction with the Markovian baths. This decaying behaviour becomes even steeper with increasing the bath temperature ($\overline{N}$). In the case of two qubits (bipartite systems) coupled to their independent reservoirs such a decaying profile is a well known phenomena and it is regarded as the ESD \cite{yu2009sudden}. We notice that a similar decaying behaviour of entanglement extends down to tripartite entangled cavities also in the presence of Markovian environments.
\subsubsection{Effect of Photon Hoping}
In order to retain entanglement for longer times in the presence of usual Markovian baths, one proposal which can be adopted is to include the effect of photon hoping among cavities. Such processes can be made more probable by placing cavities closer so that the photon wavepacket can overlap/tunnel to the other cavity. Although such a system will not be very useful for the purposes of long distance quantum communications (such as in continuous variable based quantum key distribution protocols \cite{jouguet2013experimental}), but still it may be used to store entanglement for extended times. With the inclusion of photon hoping, the system Hamiltonian as given in Eq.(\ref{H}) extends to:
\begin{equation}
\hat{H}_{new} =\hat{H}_{sys}+\hbar\xi_{12}(\hat{a}^{\dagger}_{1}\hat{a}_{2}+\hat{a}^{\dagger}_{2}\hat{a}_{1})+\hbar\xi_{23}(\hat{a}^{\dagger}_{2}\hat{a}_{3}+\hat{a}^{\dagger}_{3}\hat{a}_{2})
\end{equation}
where $\hat{H}_{sys}$ is given in Eq.(1) and the rest of terms on right side of above equation are hoping terms ($\xi_{12}$ and $\xi_{23}$ describing the hoping strength between cavity 1 and 2 and cavity 2 and 3 respectively.) In FIG.(2)c we have plotted the entanglement evolution for W-state. We notice that due to the hoping of photon, $\mathcal{N}(t)$ shows slight oscillations while decaying. Along with this it appears that the negativity vanishes at the same time (around $t=2\kappa^{-1}$) as it vanished for no hoping case (see FIG.~(2)a). But a close inspection reveals the presence of very small entanglement for later times $t>2\kappa^{-1}$ (as shown in the inset of the figure). This small entanglement exhibits the possibility of photons going back and forth between cavities due to strong hoping before being leaked out.\\
Although such a proposal can be used to save the entanglement in the system for longer times but the amount of entanglement left in the system turns out to be so small that it can hardly be useful for any practical application. Keeping in view this fact, in next section we'll investigate the effect of non-Markovian environments on the entanglement survival among cavities.
\subsection{Non-Markovian baths}
There are many quantum information protocols that however, require the sustainibility of considerable amount of entanglement for longer times. Clearly coupling of quantum systems with Markovian baths (even with the possibility of photon hoping) will not support these protocols and hence novel approaches are needed for the entanglement storage for desired amount of time. In order to address this problem, in recent years the method of entanglement control by using quantum feedback networks is developed \cite{mancini2005towards}. We on the contrary here propose to exploit the the technique of reservoir engineering through the utilization of non-Markovian baths for this purpose. \\
We discuss the dynamical behavior of tripartite entanglement, when cavities are coupled to non-Markovian baths of different Lorentzian and ohmic type spectral profiles. Here we'll first consider a relatively simple situation when the temperature of all baths is set to zero. Later on  (in section 3.3) we'll also consider the effect of finite temperatures. The main aim here will not only to compare the effect of Markovianity and non-Markovianity on entanglement but also to analyze which type of spectral density of the non-Markovian baths can surpass other types of densities in producing entanglement which is most robust against environmental decoherence.
\subsubsection{Lorentzain type Profiles}
We'll start with the examples of non-Markovian baths with Lorentzian type spectral densities. For baths at zero temperatures, EME takes a simpler form with $\alpha_{i}(t) = 0$ and $\beta_{i}(t)=\int_{0}^{t}dt_{1}\int_{0}^{\infty}J_{i}(\omega)e^{i(\omega-\omega_{ci})(t-t_{1})}d\omega$. The time-dependent decay rate $\kappa_{i}(t)$ is related to the time-dependent function $\beta_{i}(t)$ through the relation $\kappa_{i}(t) = 2{\rm Re}[\beta_{i}(t)]$. For the example of single Lorentzian bath (characterized by density $J_{SL}(\omega)$), time-dependent decay rate takes the form:
\begin{equation}
\kappa^{SL}_{i}(t)= \Bigg[\frac{\alpha_{i}\Gamma^{2}_{i}}{\delta_{i}^{2}+\Gamma^{2}_{i}}\Bigg]\Bigg[1-e^{-\Gamma_{i} t}\Bigg\lbrace cos(\delta_{i} t)-\frac{\delta_{i}}{\Gamma_{i}}sin(\delta_{i} t)\Bigg\rbrace\Bigg],
\end{equation}
whereas for double and band-gap Lorentzians we'll consider two single Lorentzian decay rates $\kappa^{SL1}_{i}(t)$ and $\kappa^{SL2}_{i}(t)$ with different widths of distribution ($\Gamma_{1}$ and $\Gamma_{2}$) which are weighted appropriately and added/subtracted to give the following time-dependent decay rates:
\begin{subequations}
\begin{align}
\kappa^{DL}_{i}(t)= W_{D1}\kappa^{SL1}_{i}(t)+W_{D2}\kappa^{SL2}_{i}(t)
\end{align}\\
\vspace{-13mm}
\begin{align}
\kappa^{BL}_{i}(t)= W_{B1}\kappa^{SL1}_{i}(t)-W_{B2}\kappa^{SL2}_{i}(t)
\end{align}\\
 \end{subequations}
 $\delta_{i} = (\omega_{ci}-\omega_{bc})$ is the detunning frequency and for simplicity we have assumed that all time-dependent decay rates are the same i.e. $\kappa_{1}(t)=\kappa_{2}(t)=\kappa_{3}(t)\equiv\kappa(t)$.\\
 
\textbf{$\bullet$ On resonance case:} With these time-dependent functions, we numerically solve the equations of motion of the density matrix elements (as outlined in Appendix A) and then the $\mathcal{N}(t)$ is calculated by applying the PPT criterion on $\rho_{s}(t)$. Similar to Markovian bath case, we notice that only one of the eigenvalues in the partial transposed density matrix (we call it $\lambda_{-}$) exhibit negative character as required by the definition of negativity, thus leading to: $\mathcal{N}(t)= max(0,\hspace{2mm}2\lambda_{-}(t))$. In Fig. 3(a),-(c) we have plotted the dynamics of tripartite entanglement when cavities are initially prepared in W and GHZ states respectively. 
All three examples (single, -double and -band-gap Lorentzians) are shown along with Markovian case. Here firstly we have presented the results for on resonance case i.e. $\delta=0$.\\

In Fig.~3(a),-(c), we notice that utilizing a non-Markovian bath of any type of Lorentzian spectral profile results in somewhat slower decay (and hence longer survival) of entanglement compared to the usual Markovian bath case. Double Lorentzian bath makes this behavior most pronounced. GHZ state has initially (at $t=0$) slightly more entanglement than W state and that difference prevails throughout the plot. In W and GHZ states, system start off in maximally entangled tripartite states, but reaching at $t\simeq 8, 5, 4 $ and $3$ (units of time) for double, -single, -band-gap Lorentzian and flat Markovian baths respectively, entanglement completely dies out due to photon leakage. \\
The different decaying slopes of $\mathcal{N}(t)$ in all curves, can be explained by plotting the corresponding time-dependent decay rate (as shown in Fig. 3(b)). Since the decay rates have time-dependence therefore we calculate the average decay rate over entire time interval (i.e. $\overline{\kappa}(t)=\frac{1}{\Delta T}\int_{t}^{t+\Delta T}\kappa(t^{'})dt^{'}$ whereas in present case $\Delta T=10\omega^{-1}_{c}$). We find that for double -single and band-gap Lorentzians the average decay rate turns out to be 0.42, 0.74 and 1.37 $\omega^{-1}_{c}$ respectively. These time-dependent rates represent effective decay seen by the cavities and it  explains why for example the entanglement decays most slowly (fastly) in double (band-gap) Lorentzian case compared to other situations.\\
\begin{figure*}[t]
\centering
  \subfloat{%
    \includegraphics[width=5.4cm,height=4.4cm]{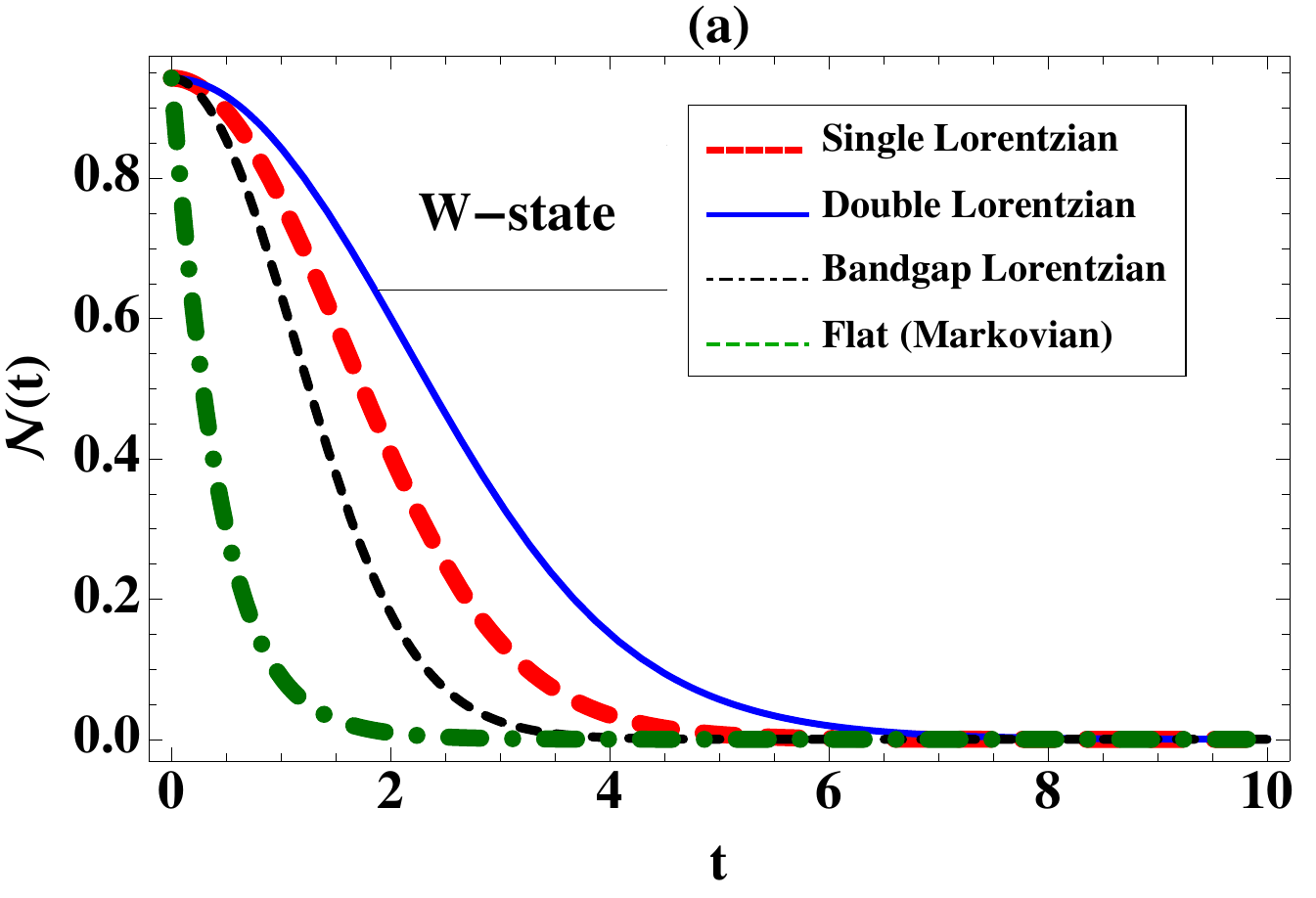}}
  \subfloat{%
    \includegraphics[width=5.4cm,height=4.4cm]{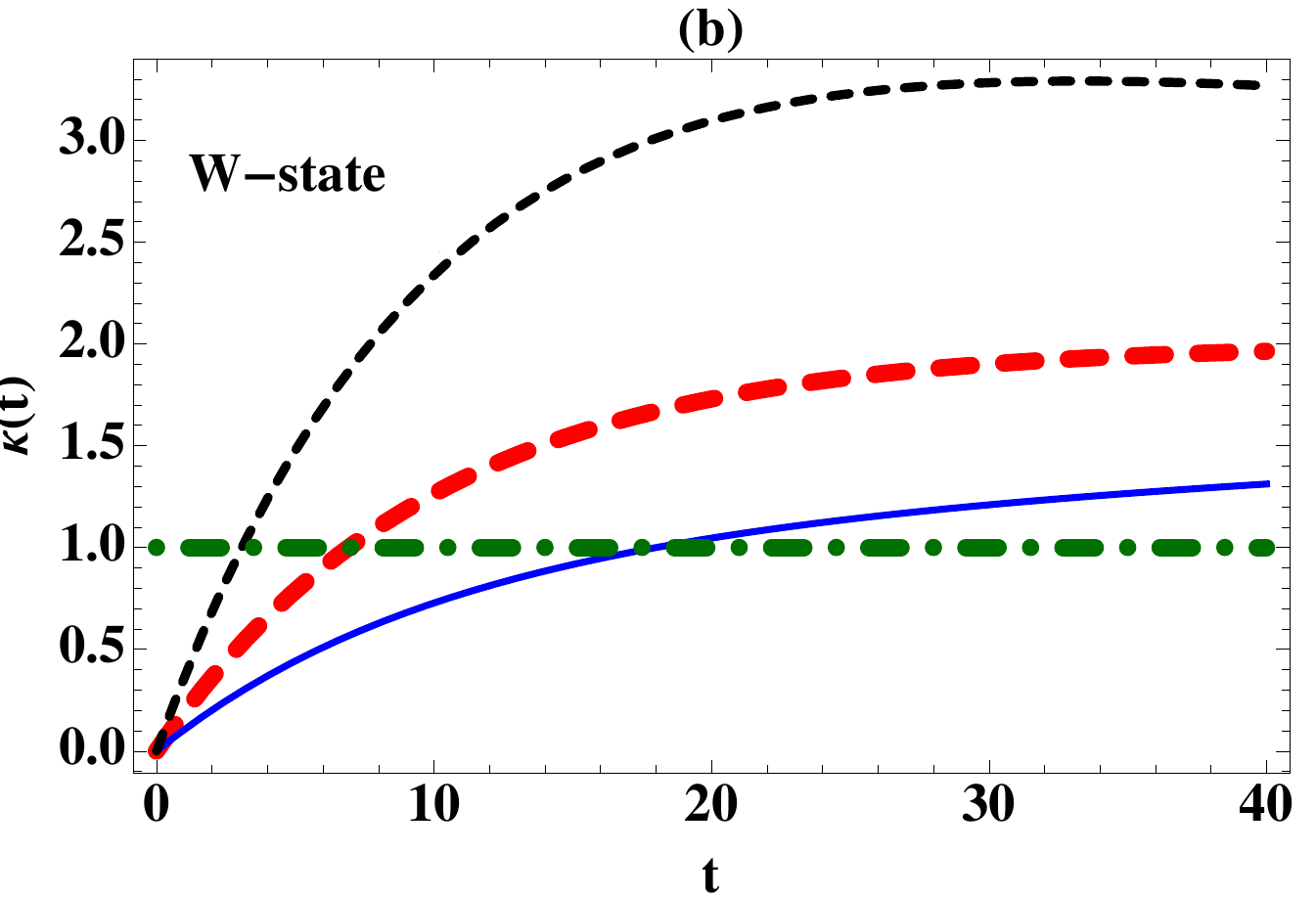}}
    \subfloat{%
    \includegraphics[width=5.4cm,height=4.4cm]{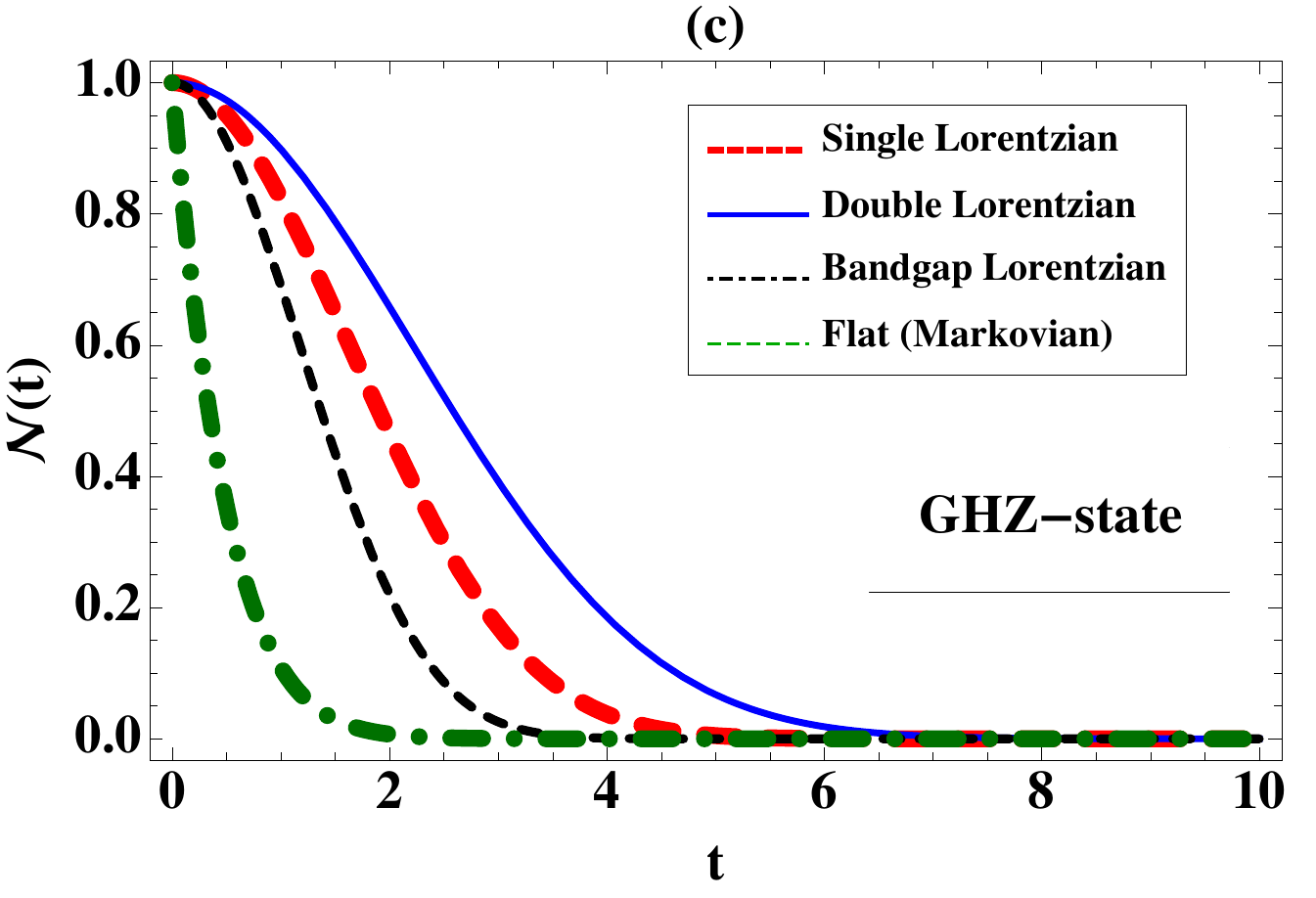}}
    \captionsetup{
  format=plain,
  margin=1em,                          
  justification=raggedright,
  singlelinecheck=false
}
  \caption{Temporal profile of entanglement (negativity) among three optical cavities when system start off in a (a) W state (c) GHZ state, for different Lorentzian non-Markovian baths as well as for flat Markovian bath. 
 In all plots $\delta=\omega_{c}-\omega_{bc}=0$ and for single Lorentzian bath $\alpha_{L} = 2, \Gamma=0.1\omega_{c}$. For double and band-gap Lorentzian common parameters are: $\alpha_{L1} = \alpha_{L2} =2, \Gamma_{1}=0.1\omega_{c},\Gamma_{2}=0.01\omega_{c}$ whereas for double (band-gap) Lorentzian $W_{1}=W_{2}=1/2$ $(W_{1}=2, W_{2}=1)$ are chosen. }\label{Fig3}
\end{figure*}

\textbf{$\bullet$ Off resonance case:} In Fig.~4 we plot the negativity and time-dependent decay rates as a function of time under finite detunning case. We note that by introducing $\delta \neq 0$, entanglement start to show oscillations superimposed by an overall decay. Why entanglement shows fast and slow decay and collapse and revival against different spectral baths for off resonance case? We'll answer this question by employing the NMQJA developed by Piilo et.al in \cite{piilo2008non} (see Appendix B for a brief review).\\
{\bf Explanation of negativity's behavior based non-Markovian quantum jump approach (NMQJA)}\\
 To apply NMQJA,  we have plotted the time-dependent decay rates for different types of Lorentzian environmental models in all cases.  We note that the periodic temporal structure of decay rate translates to $\mathcal{N}(t)$. During the positive cycle of decay rate, entanglement decreases (positive jump occurs), while during the negative cycle of the decay rate, non-Markovian environment transfers (some part of) entanglement back to the system (negative jump occurs).\\
  It seems surprising though that the entanglement evolution for both W and GHZ states behave in exact same manner. In terms of NMQJA, this difference can be seen by considering for example that the system start off in a GHZ state. Say during the positive cycle, one photon is leaked by each cavity then the initial state of cavities $\ket{111}$ can be in any one of the kets $\ket{011},\ket{101},\ket{110}$. But it is also possible that during this time another photon is emitted then the system will be found in either $\ket{100}$ or in $\ket{010},\ket{001}$. Now later when negative cycle arrives two negative jumps will be needed to obtain the initial maximally entangled GHZ state. Clearly for an initial W-state only one positive jump is possible which can be compensated by one negative jump. Thus one will expect less entanglement regain for GHZ state as compared to W state upon a negative cycle of decay rate, which is ofcourse not the case (as seen in Fig.~4). This identicalness of $\mathcal{N}(t)$ behaviors in W and GHZ states can be explained by assuming that the decay rate is weak ($\lbrace\Gamma_{1},\Gamma_{2}\rbrace<\alpha$) and hence the possibility of a second photon decay is almost negligible. At a single photon leakage level thus one will observe an identical $\mathcal{N}(t)$ profile (as seen in Fig.~4).\\  
In single Lorentzian case (Fig. 4(a) and (d)) $\kappa(t)$ oscillates between 0.58 and -0.28 and hence entanglement shows clear collapses and revivals. While for double Lorentzian case (Fig. 4(b) and (e)) the amplitude of the decay decreases and remains between 0.3 and -0.2, thus $\mathcal{N}(t)$ doesn't fully dies out before the revival cycle arrives and entanglement again shows an increase. Finally we remark that the double Lorentzian bath  produces the most robust entanglement against decoherence (considerable entanglement survives till $20\kappa^{-1}$) dominating all other types of baths considered in Fig.~4. Here we would also like to point out that there has been some studies conducted in past where effect of detuning on the entanglement preservation between two qubits has been investigated \cite{bellomo2008entanglement, xiao2009robust}. It is interesting to note that the detuning turns out to be a useful parameter for entanglement storage for the tripartite system under study as well.
\begin{figure*}[t]
\centering
  \subfloat{%
    \includegraphics[width=5.2cm,height=4cm]{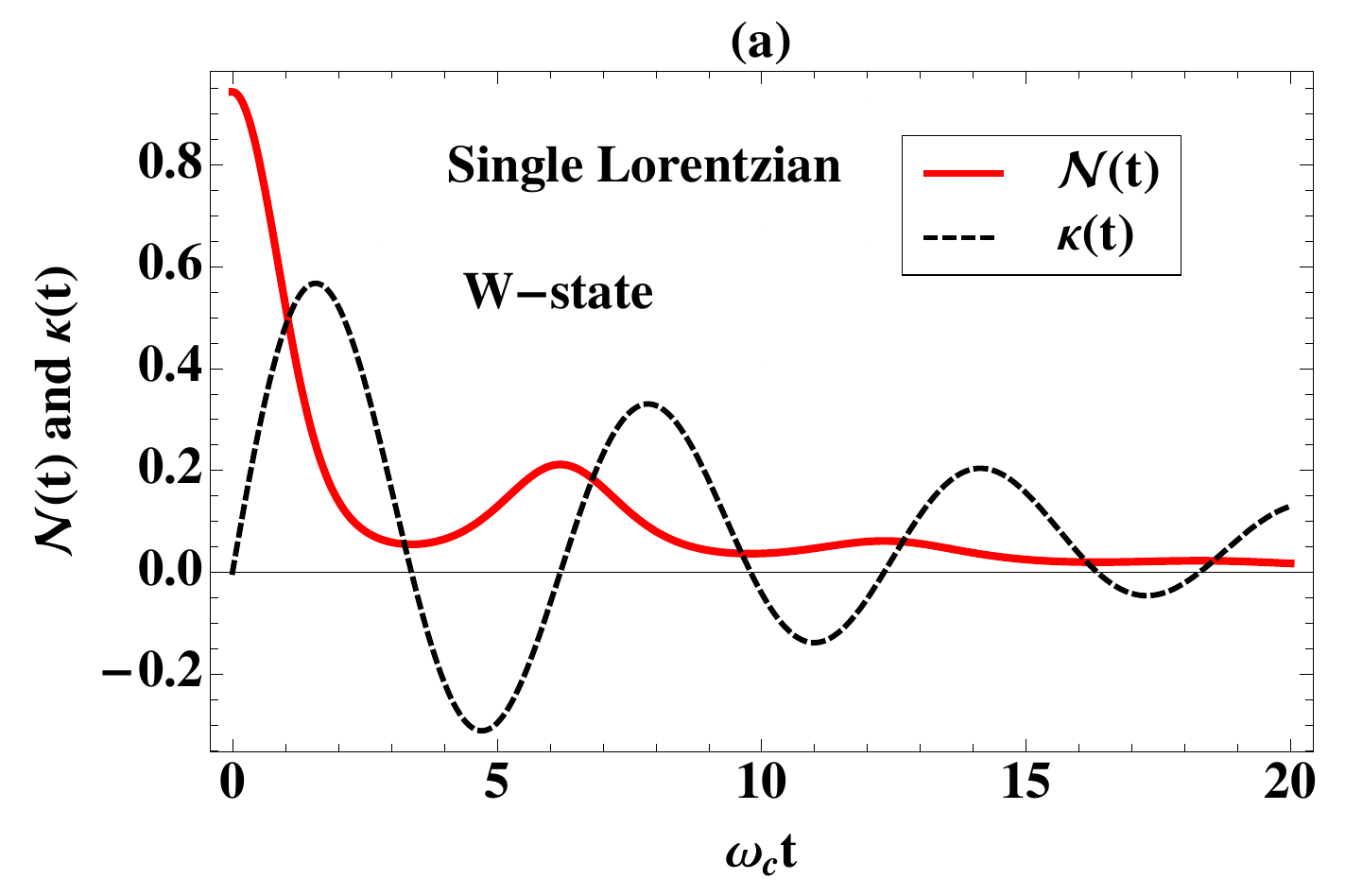}}
  \subfloat{%
    \includegraphics[width=5.2cm,height=4cm]{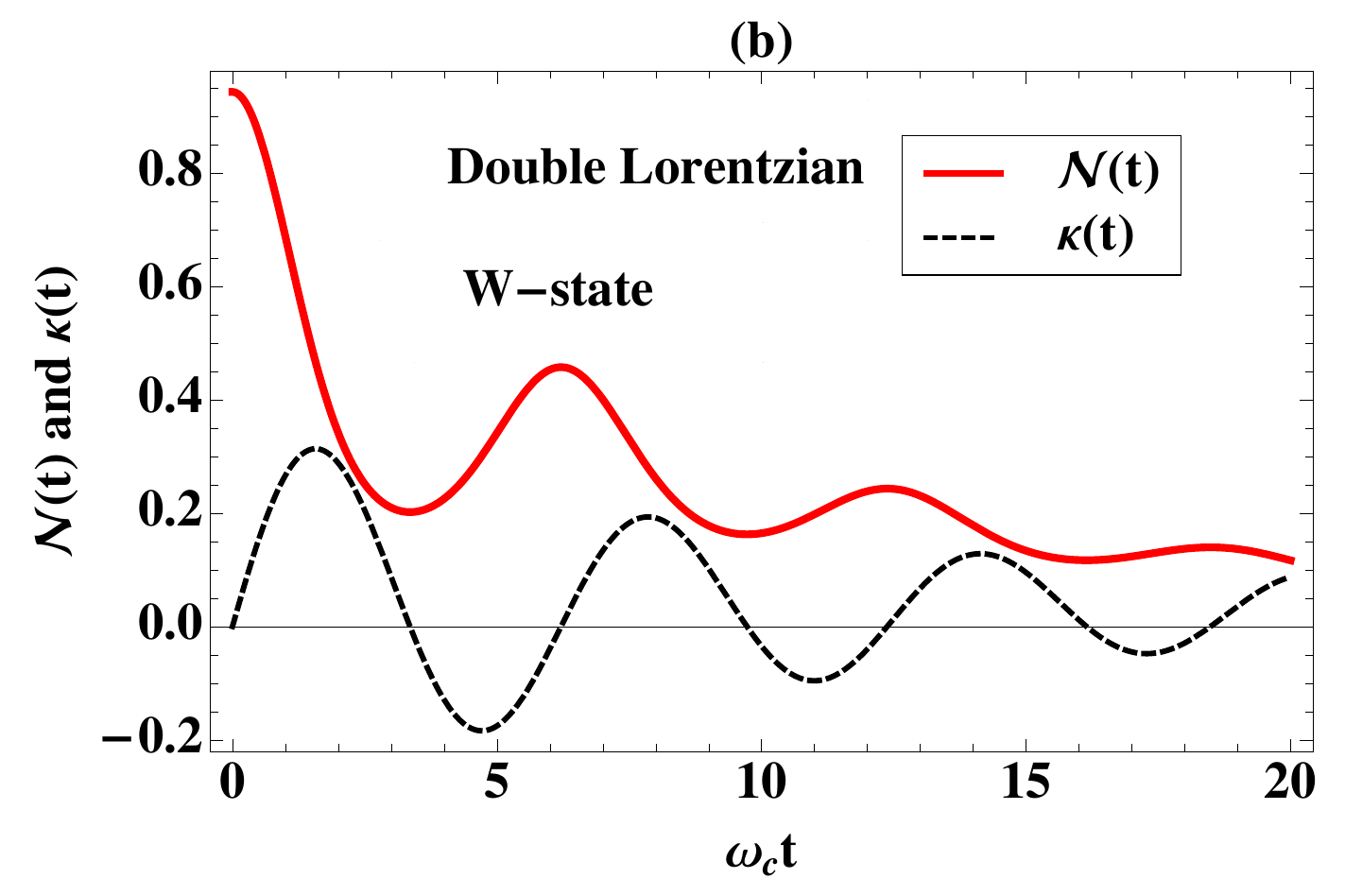}}
  \subfloat{%
    \includegraphics[width=5.2cm,height=4cm]{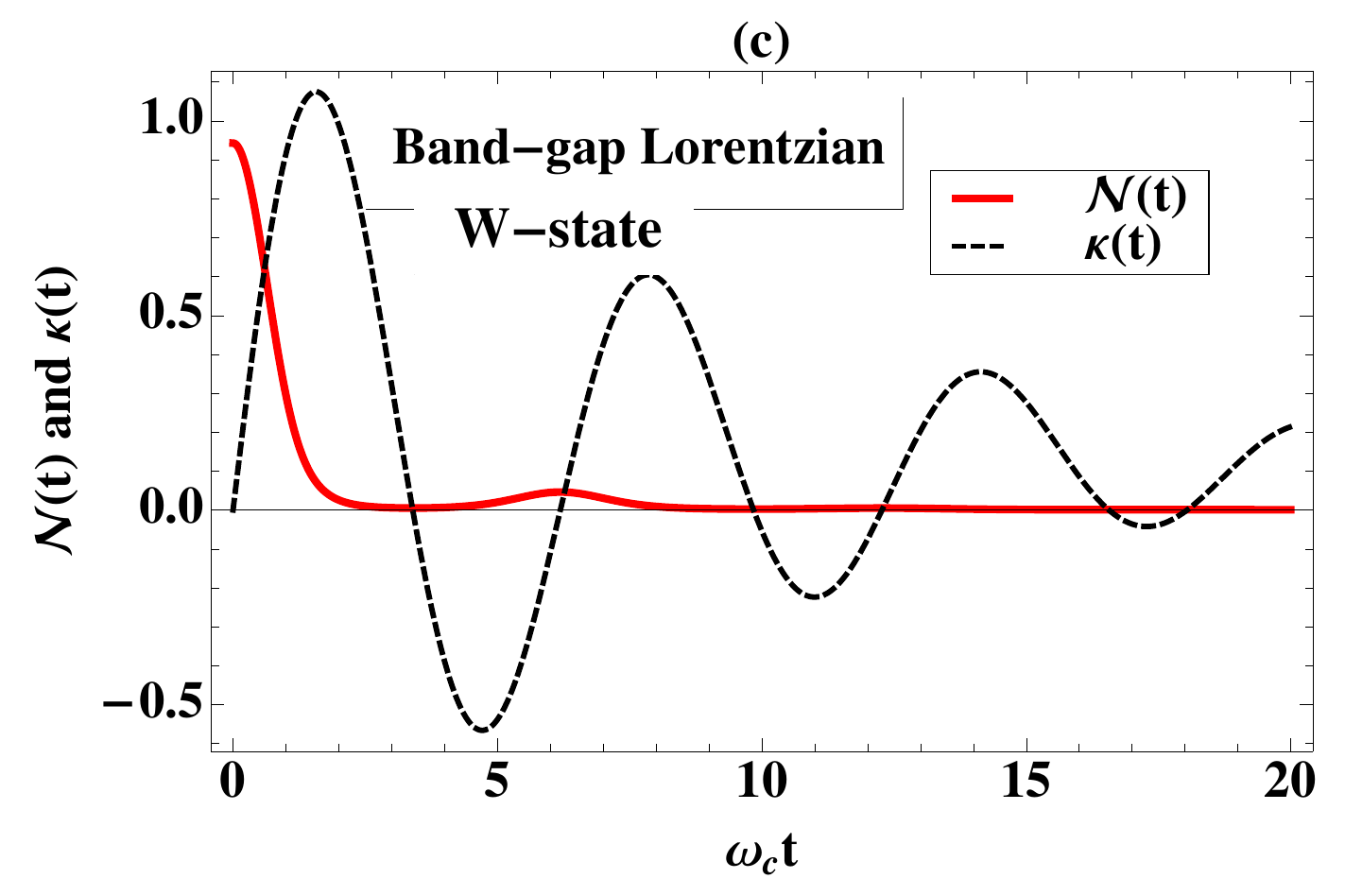}}\\
  \subfloat{%
    \includegraphics[width=5.2cm,height=4cm]{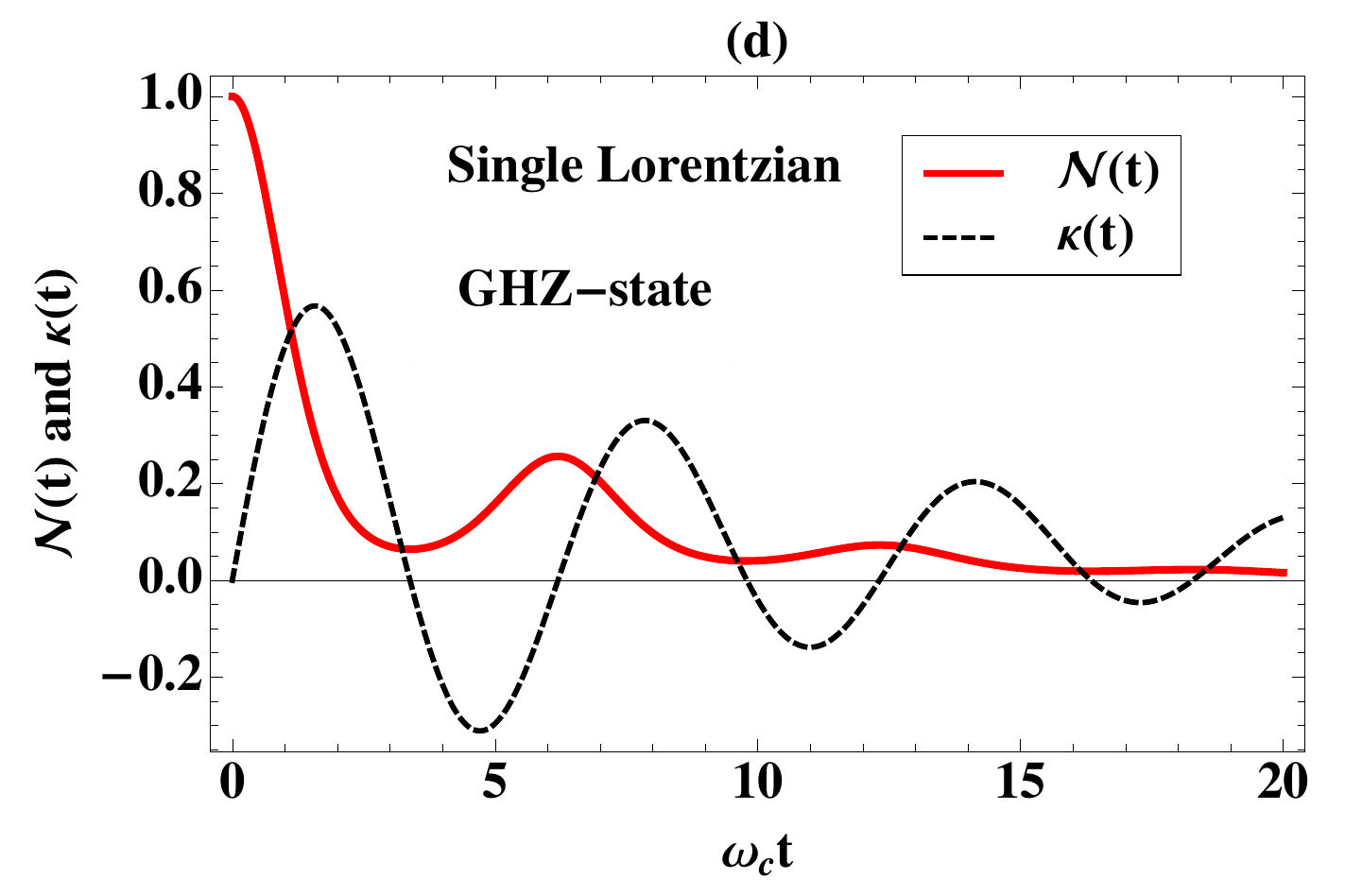}}
  \subfloat{%
    \includegraphics[width=5.2cm,height=4cm]{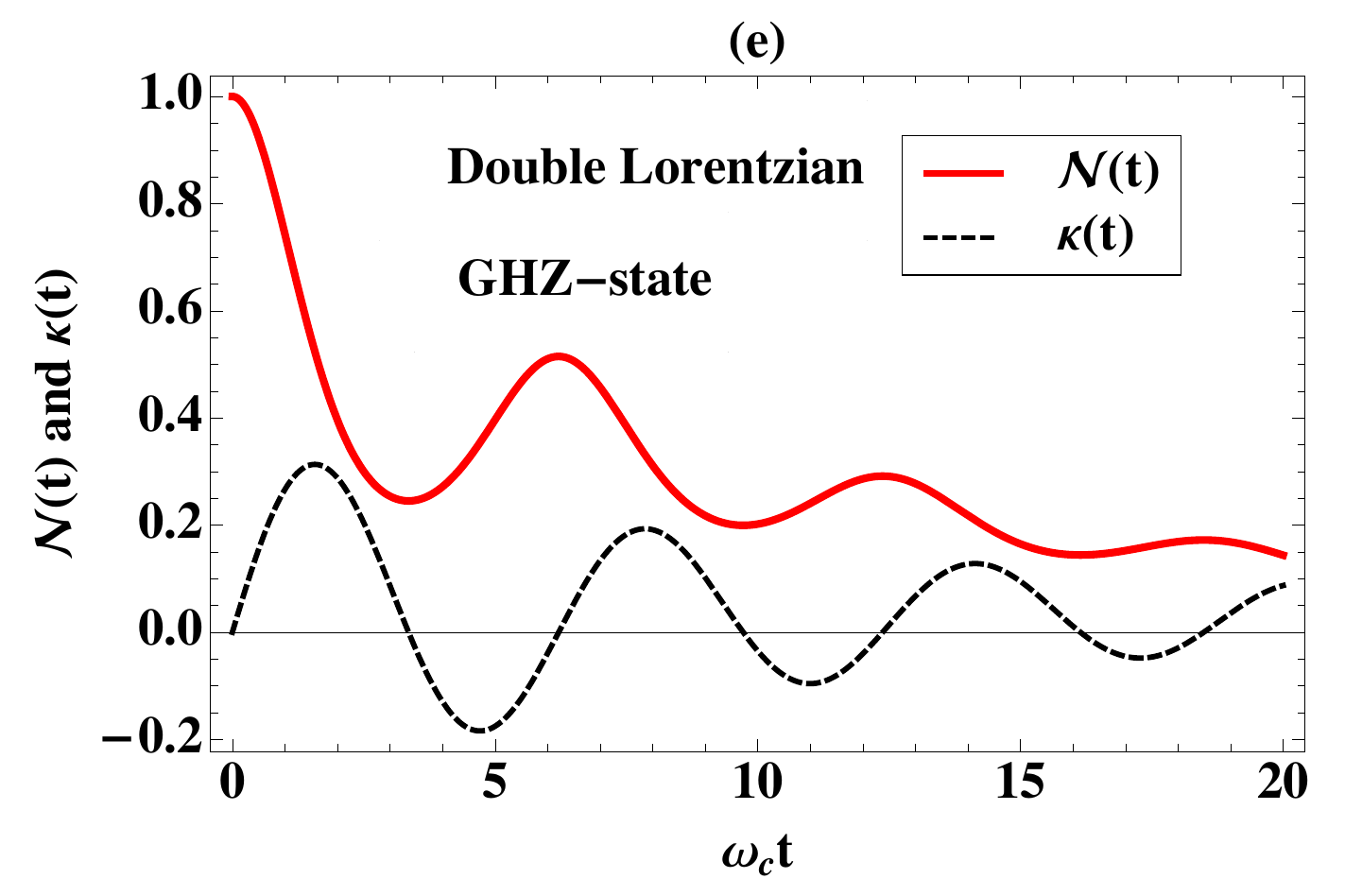}}
  \subfloat{%
    \includegraphics[width=5.2cm,height=4cm]{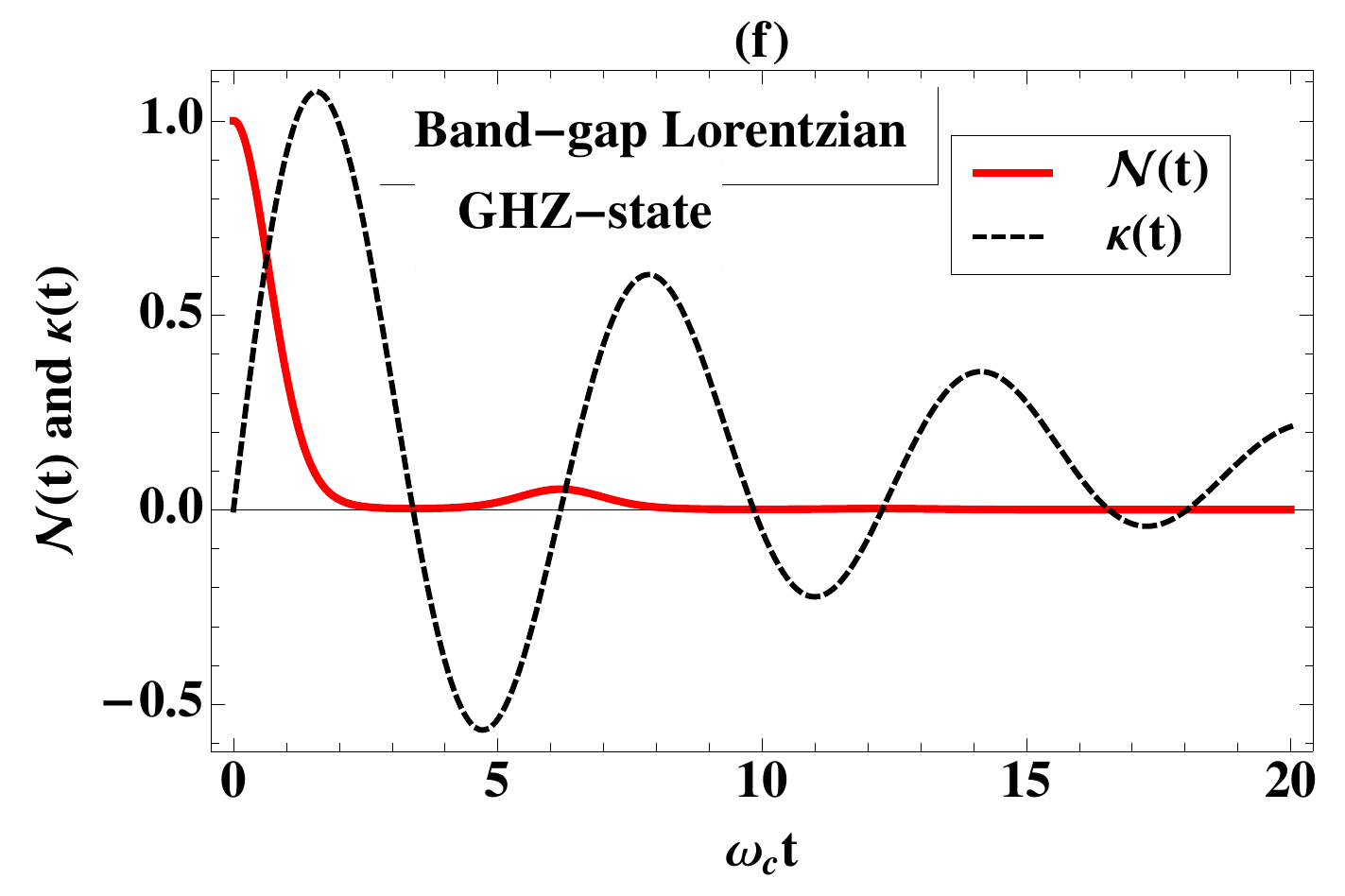}}
    \captionsetup{
  format=plain,
  margin=1em,                          
  justification=raggedright,
  singlelinecheck=false
}
  \caption{Entanglement and time-dependent decay rates plotted versus dimensionless time $\omega_{c}t$ for Lorentzian models for an initial (a),(b),(c) W and (d),(e),(f) GHZ state inclusing detunning $\delta = 1\omega_{c}$. Other parameters used (in all parts of the figure) are: $\alpha=\alpha_{1}=\alpha_{1}=6, \Gamma=\Gamma_{1}=0.1\omega_{c}, \Gamma_{2}=0.01\omega_{c}$. Weightage factors ($W_{1}, W_{2}$) are the same as chosen in Fig.~3. Note that due to non-zero detunning, entanglement shows oscillatory character, with a possibility of collapse and revival. }\label{Fig4}  
\end{figure*}

\begin{figure*}[t]
\centering
  \subfloat{%
    \includegraphics[width=5.4cm,height=4.4cm]{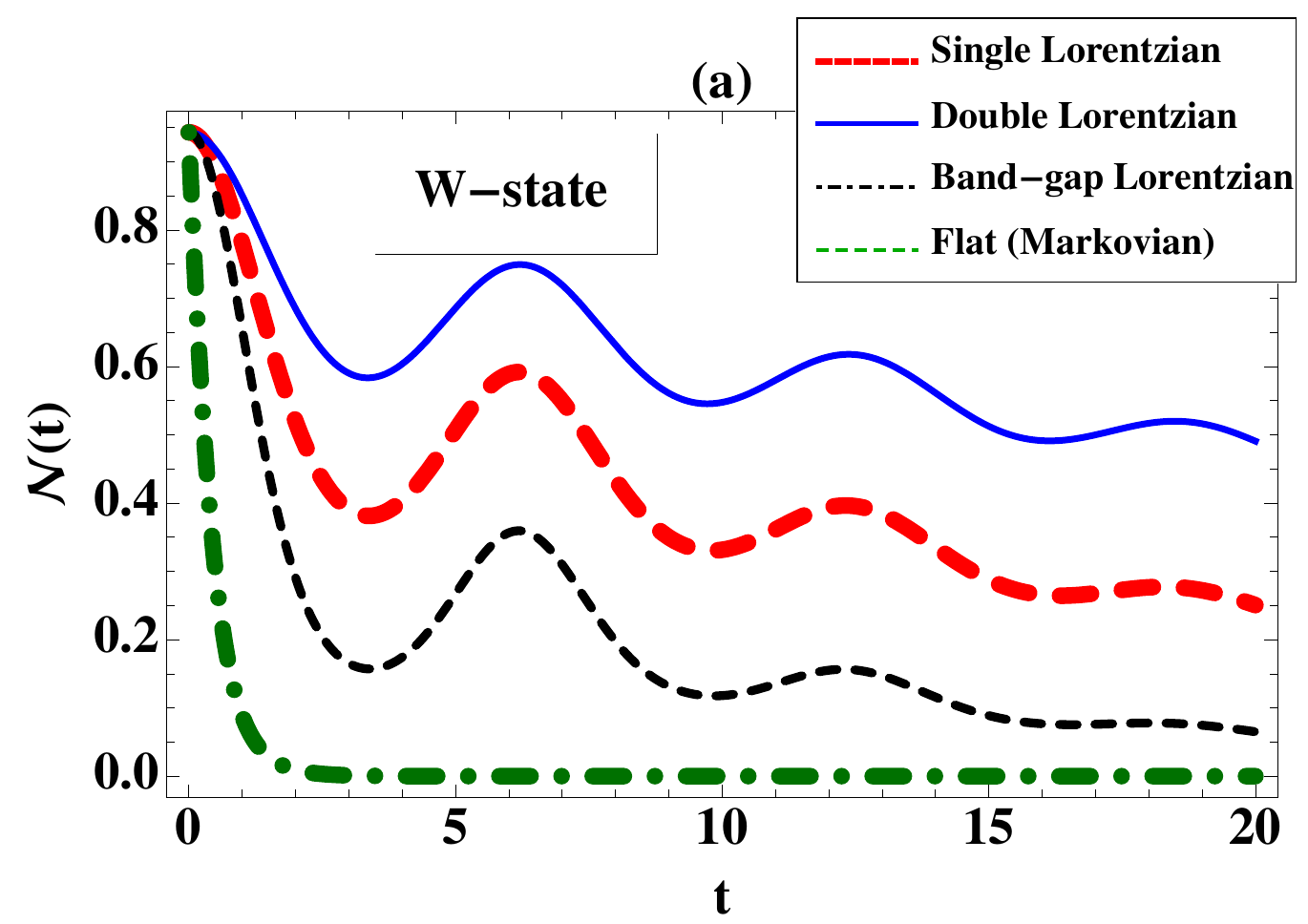}}
  \subfloat{%
    \includegraphics[width=5.4cm,height=4.4cm]{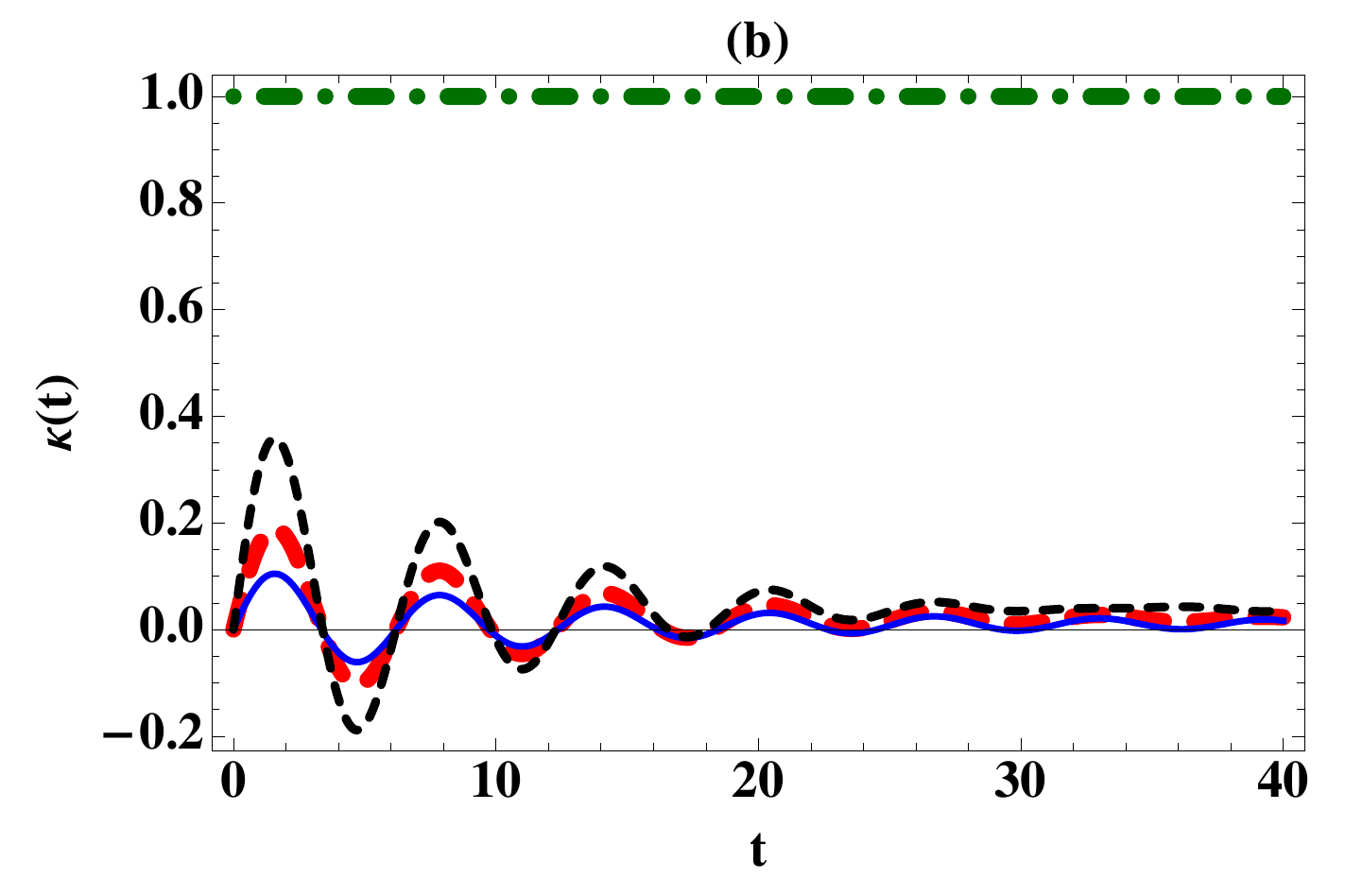}}
    \subfloat{%
    \includegraphics[width=5.4cm,height=4.4cm]{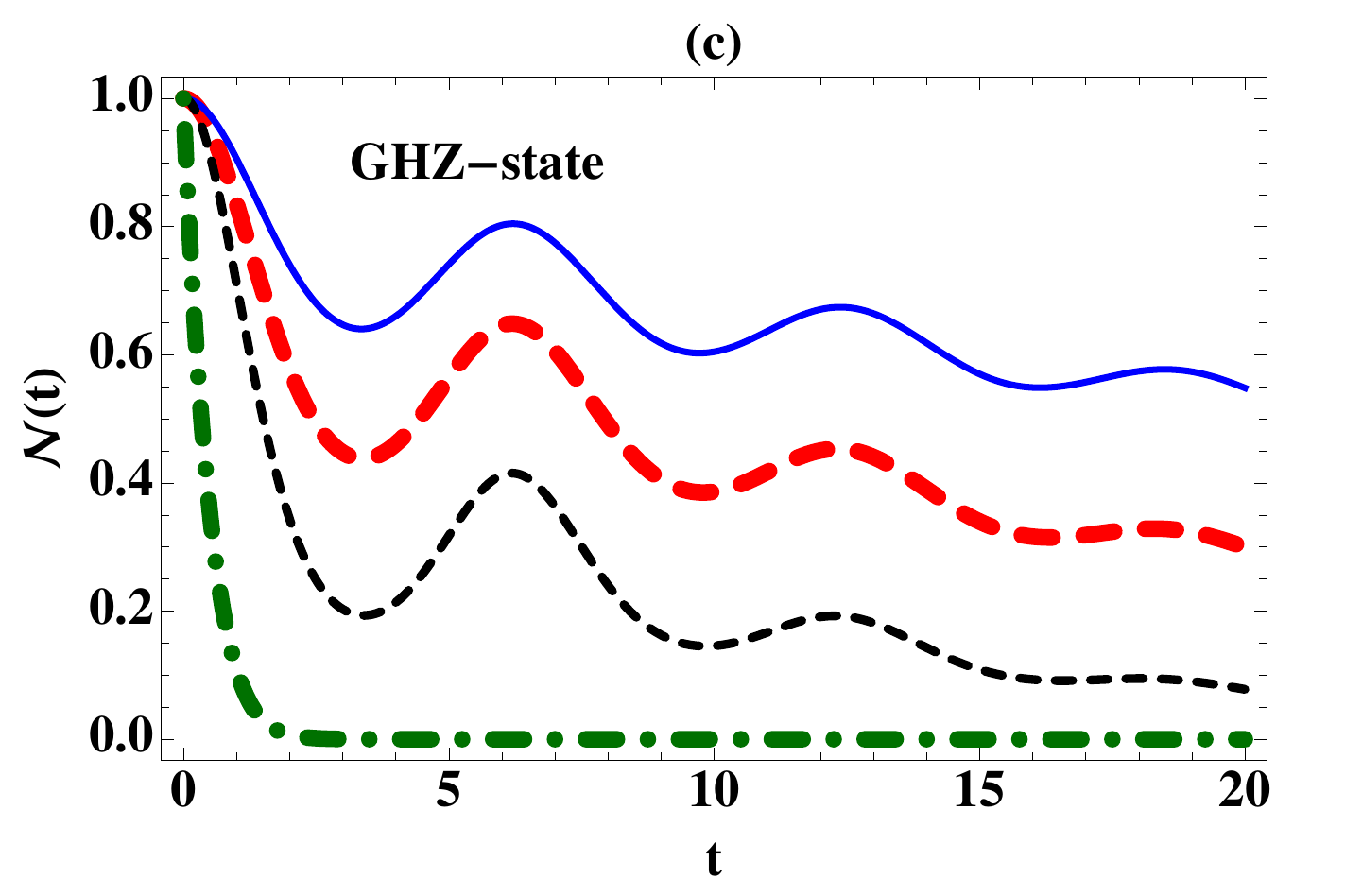}}
    \captionsetup{
  format=plain,
  margin=1em,                          
  justification=raggedright,
  singlelinecheck=false
}
  \caption{Effect of reducing the dimensionless coupling constant $\alpha_{L}, \alpha_{L1}, \alpha_{L2}$ on non-Markovian tripartite entanglement dynamics. In all plots  $\alpha_{L}=\alpha_{L1}=\alpha_{L2}=2$ is considered and time axis is measured in $\omega^{-1}_{c}$ units, rest of the parameters are the same as used in Fig.~4. Small amplitudes and oscillations in $\kappa(t)$ explains the higher and oscillatory negativity. }\label{Fig5}
\end{figure*}
\textbf{$\bullet$ Effect of decreasing coupling constants:} In order to further elevate the tripartite entanglement for initial times, we varied different parameters involved in the dynamics of negativity. We identified that by decreasing the strength of dimensionless coupling constants ($\alpha_{L}, \alpha_{L1}, \alpha_{L2}$) this issue can be resolved. Fig.~5 exhibits this effect, where we have selected  $\alpha_{L}(= \alpha_{L1}=\alpha_{L2}=2$), three times smaller then the values considered in Fig.~4. As a result, non-Markovian entanglement now shows much higher values than the one found in Fig.~4 for initial times in particular and for entire time range in general. From the perspective of NMQJA this happens due to smaller amplitudes of the decay rate $\kappa(t)$ compared to the ones seen in Fig.~4. Alternately in the pseudomode picture of non-Markovian baths this behavior can be explained by arguing that with the consideration of smaller coupling constants the interaction of the bath cavity mode with the outside Markovian environment decreases as a result photon remains inside the system cavity and $\mathcal{N}(t)$ shows longer survival.
\\
\begin{figure*}[t]
\centering
  \subfloat{%
    \includegraphics[width=5.4cm,height=4.4cm]{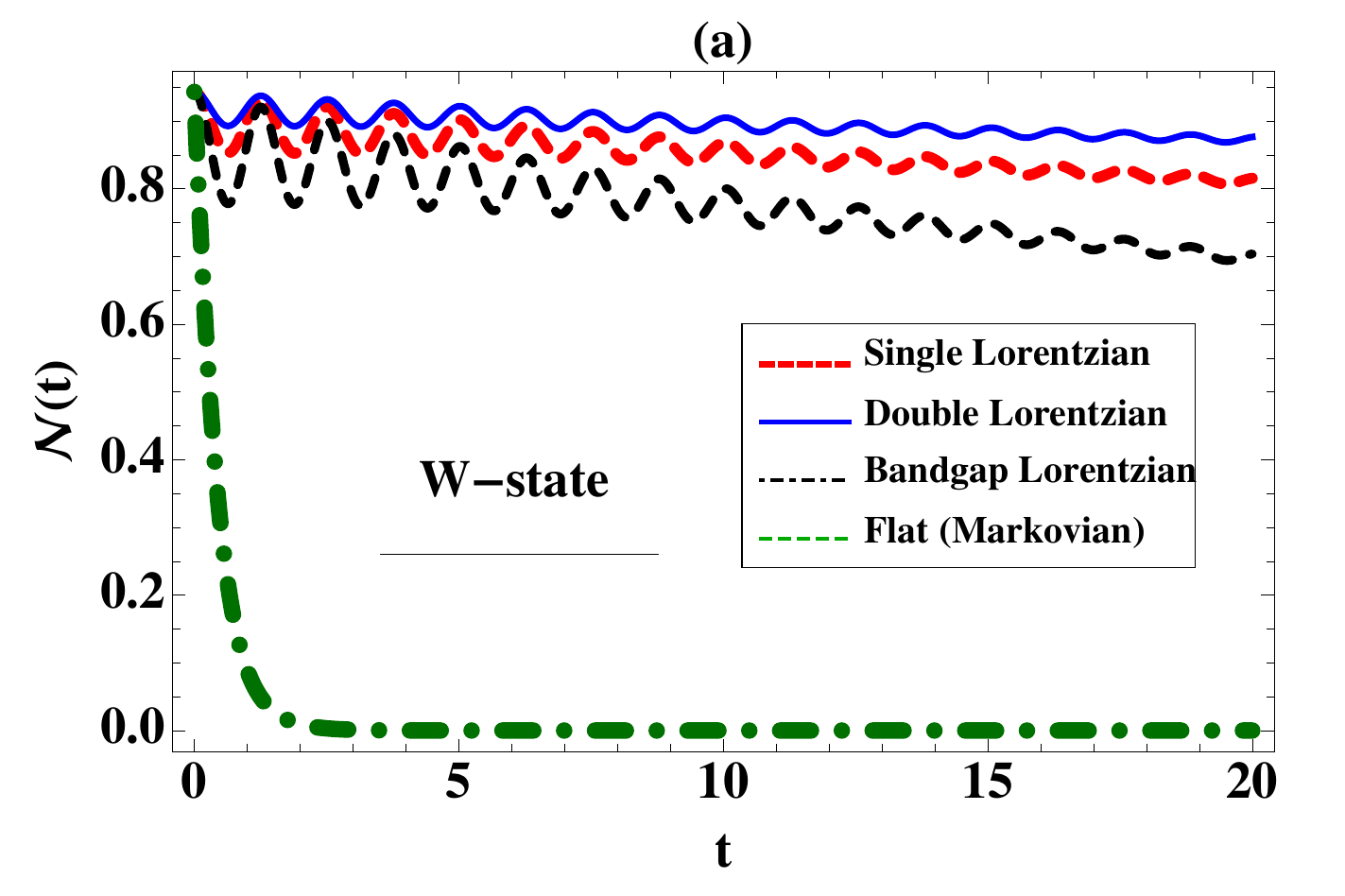}}
  \subfloat{%
    \includegraphics[width=5.4cm,height=4.4cm]{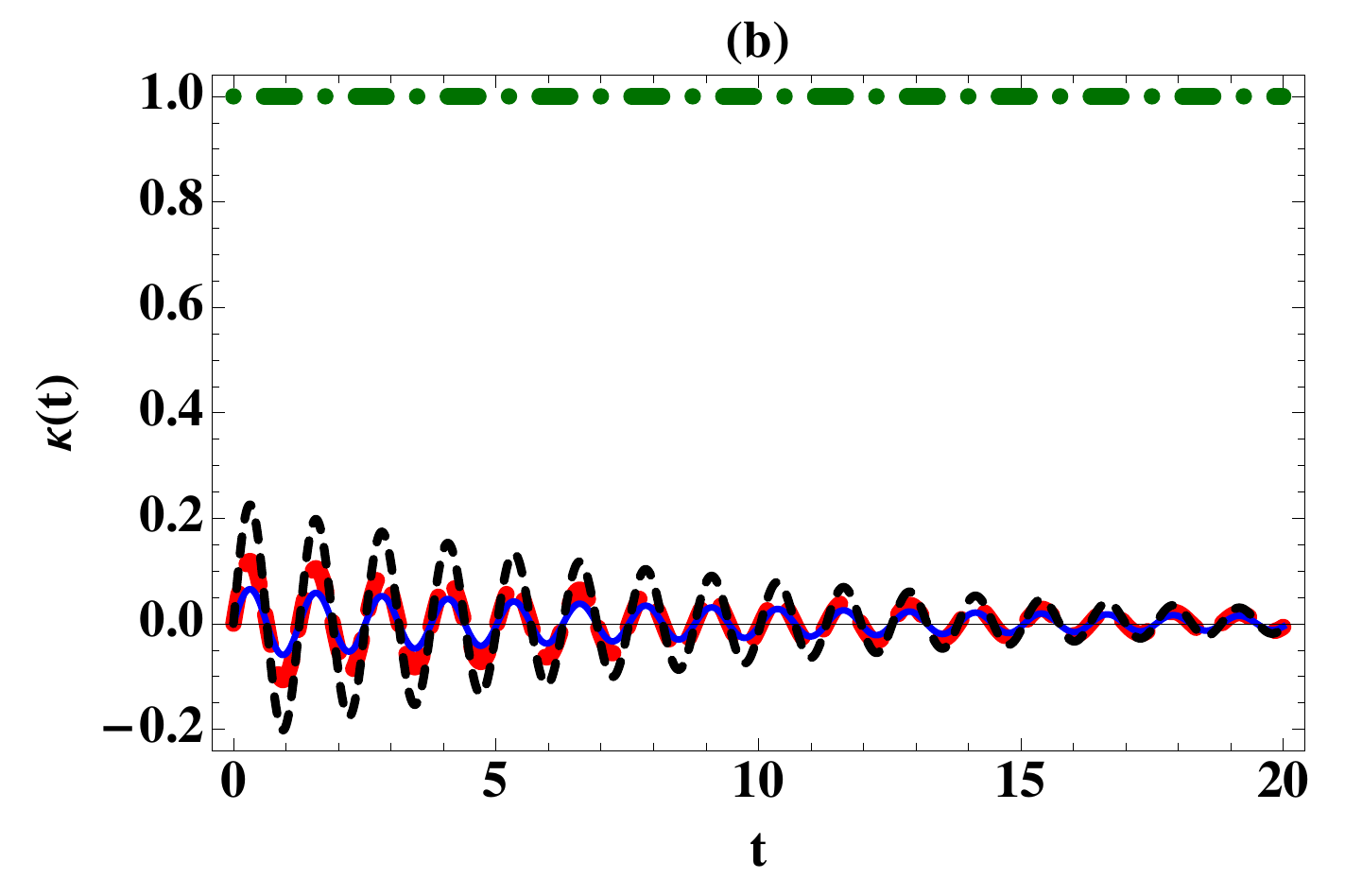}}
    \subfloat{%
    \includegraphics[width=5.4cm,height=4.4cm]{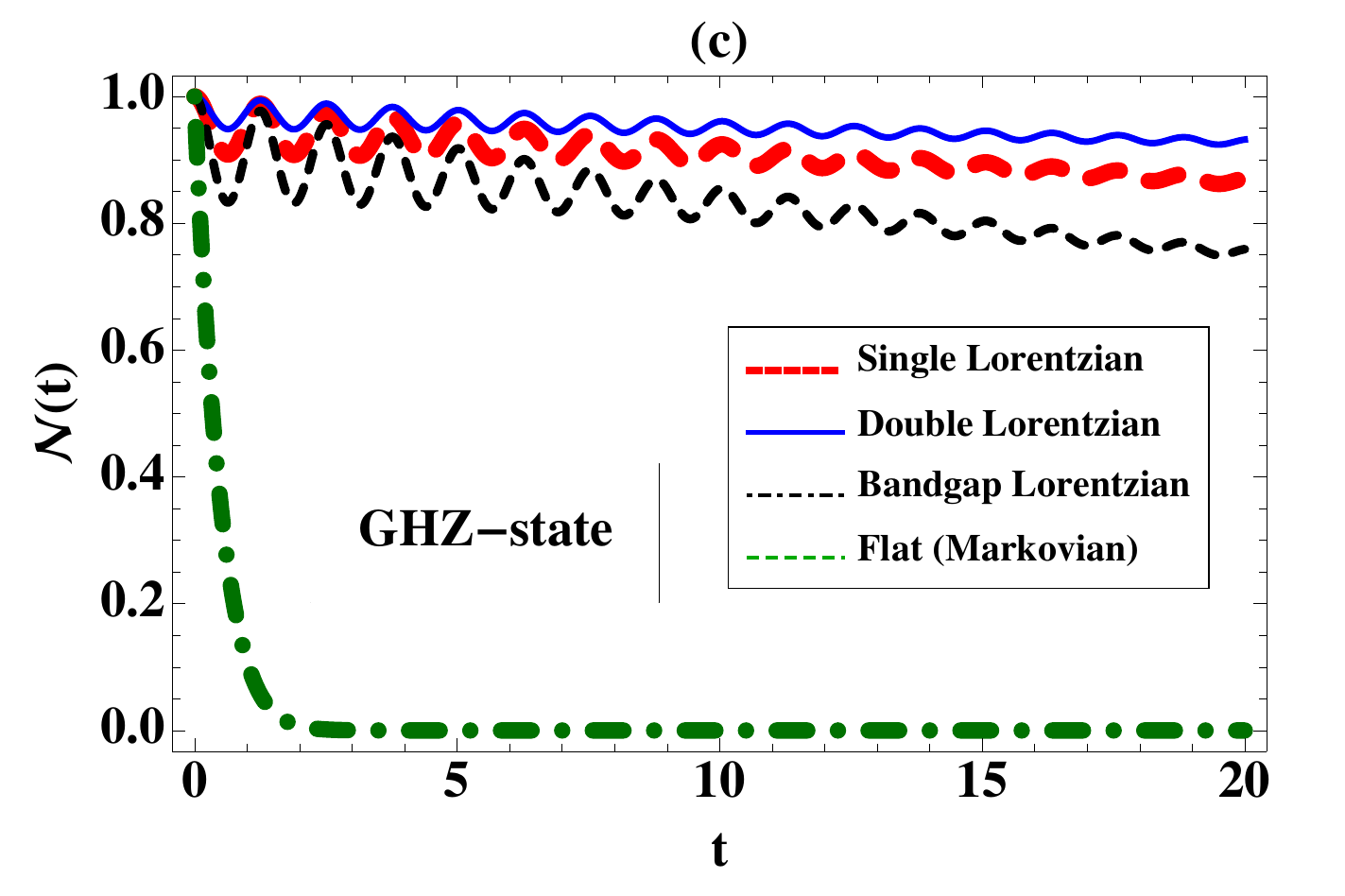}}
    \captionsetup{
  format=plain,
  margin=1em,                          
  justification=raggedright,
  singlelinecheck=false
}
  \caption{Tripartite-cavity entanglement variation in time for far detunned case $\delta=5\omega_{c}$ (rest of the parameters are the same as used in Fig.~4 and with time measured in $\omega^{-1}_{c}$ units). Exploiting the off-resonance photonic decays, we notice that $\mathcal{N}(t)$ shows an extremely robust behavior against environmnetal interactions. Substantial entanglement survives upto $t\sim 20\omega^{-1}_{c}$ in all Lorentzian cases which is due to very small oscillation amplitude gained by $\kappa(t)$ (as shown in Fig.~6(b)) in accordance to NMQJA. }\label{Fig6}
\end{figure*}
\textbf{$\bullet$ Far detunned case:} Another range of parameter that we found relevant is the regime where bath cavity is considered to be far detunned from the system cavity. We find that (as shown in Fig.~6) the consequences of considering far-off resonant Lorentzian baths ($\delta=5\omega_{c}$) is substantial. In all types of Lorentzian baths entanglement shows tiny oscillations with no more than $20\%$ of decay to the initial value upto $t=20\omega^{-1}_{c}$. Especially for double Lorentzian case, $\mathcal{N}(t)$ remains within $10\%$ of its initial value, hence showing very small deterioration. From the perspective of NMQJA, this behavior is attributed to the smallest amplitude gained compared to cases considered in Fig.~4 and 5. Thus we concluded that the far detunned Lorentzian baths produces an almost decoherence free evolution of entanglement among cavities irrespective of the initial entangled state. 
\subsubsection{Ohmic type Profiles}
Next we address the non-Markovian baths with ohmic type of spectral densities. Such type of spectral density corresponds to classical models of velocity-dependent friction forces and hence are termed as the ohmic functions. For the density function given in Eq.5(d), the time-dependent decay rate in present case turns out to be:
\begin{equation}
\kappa(t)=\frac{\alpha}{2}\Bigg( \frac{1-cos[(s-1)tan^{-1}(\omega_{cut}t)]G(s-1)}{(1+\omega^{2}_{cut}t^{2})^{\frac{s-1}{2}}}\Bigg)
\end{equation}
where $G(s-1)$ is the Euler gamma function. To investigate the influence of these types of non-Markovian models on tripartite entanglement among cavities, in Fig.~7 we have plotted the negativity dynamics and time-dependent decay rates for sub-ohmic, ohmic and super-ohmic baths. For both (W and GHZ) type of initially entangled cavities, we find that due to smaller cut-off frequencies super-ohmic baths work very well compared to the Markovian and other types of ohmic baths to store entanglement (as shown in Fig.7(a),-(b) where considerable amount of $\mathcal{N}(t)$ survives until $t\sim 10$). Utilization of sub-ohmic and ohmic baths on the other hand turns out to be not very useful. In ohmic case although the negativity remains slightly higher than the Markovian case entanglement, but it dies out almost at the same time ( $t\sim 2$) and after that due to the absence of negative cycle in $\kappa(t)$ no revival occurs. Most strikingly, the sub-ohmic non-Markovian baths works even worse than the Markovian baths for the purposes of sustaining entanglement and $\mathcal{N}(t)$ vanishes almost at $t\sim 0.5$. Again time average decay can give an effective rate seen by each cavity which is responsible for smaller/faster negativity decay.
\begin{figure*}[t]
\centering
  \subfloat{%
    \includegraphics[width=5.4cm,height=4.4cm]{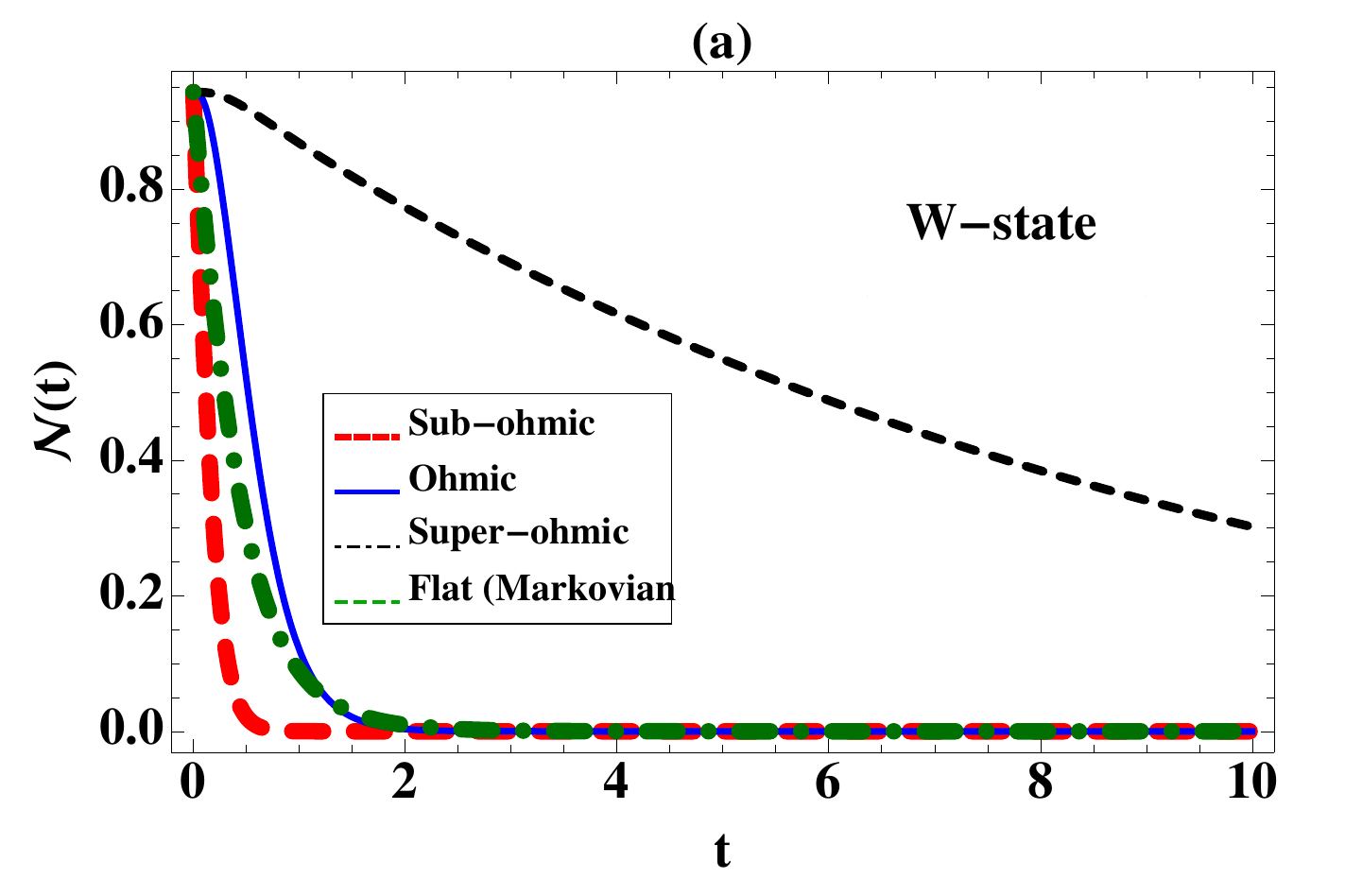}}
  \subfloat{%
    \includegraphics[width=5.4cm,height=4.4cm]{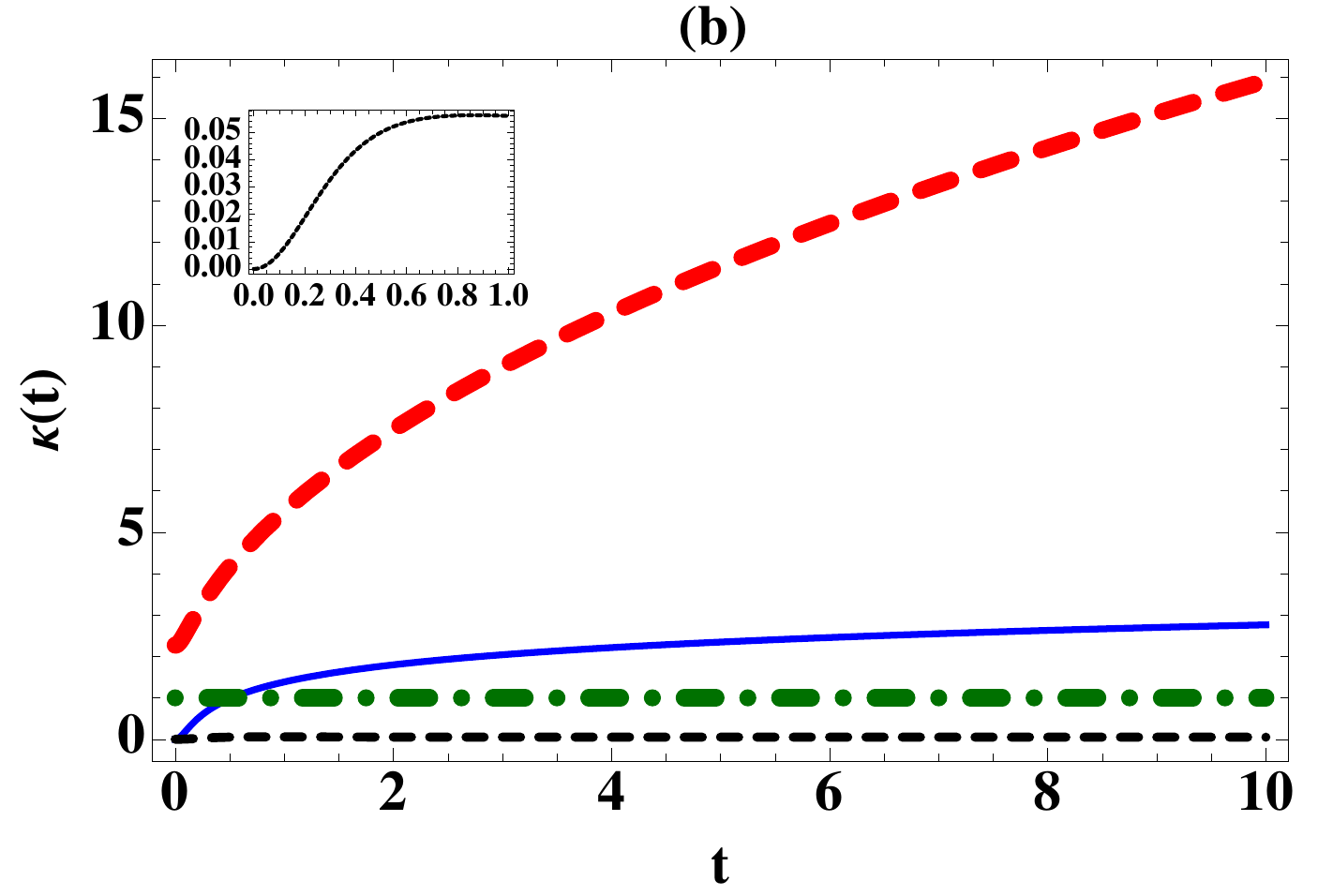}}
    \subfloat{%
    \includegraphics[width=5.4cm,height=4.4cm]{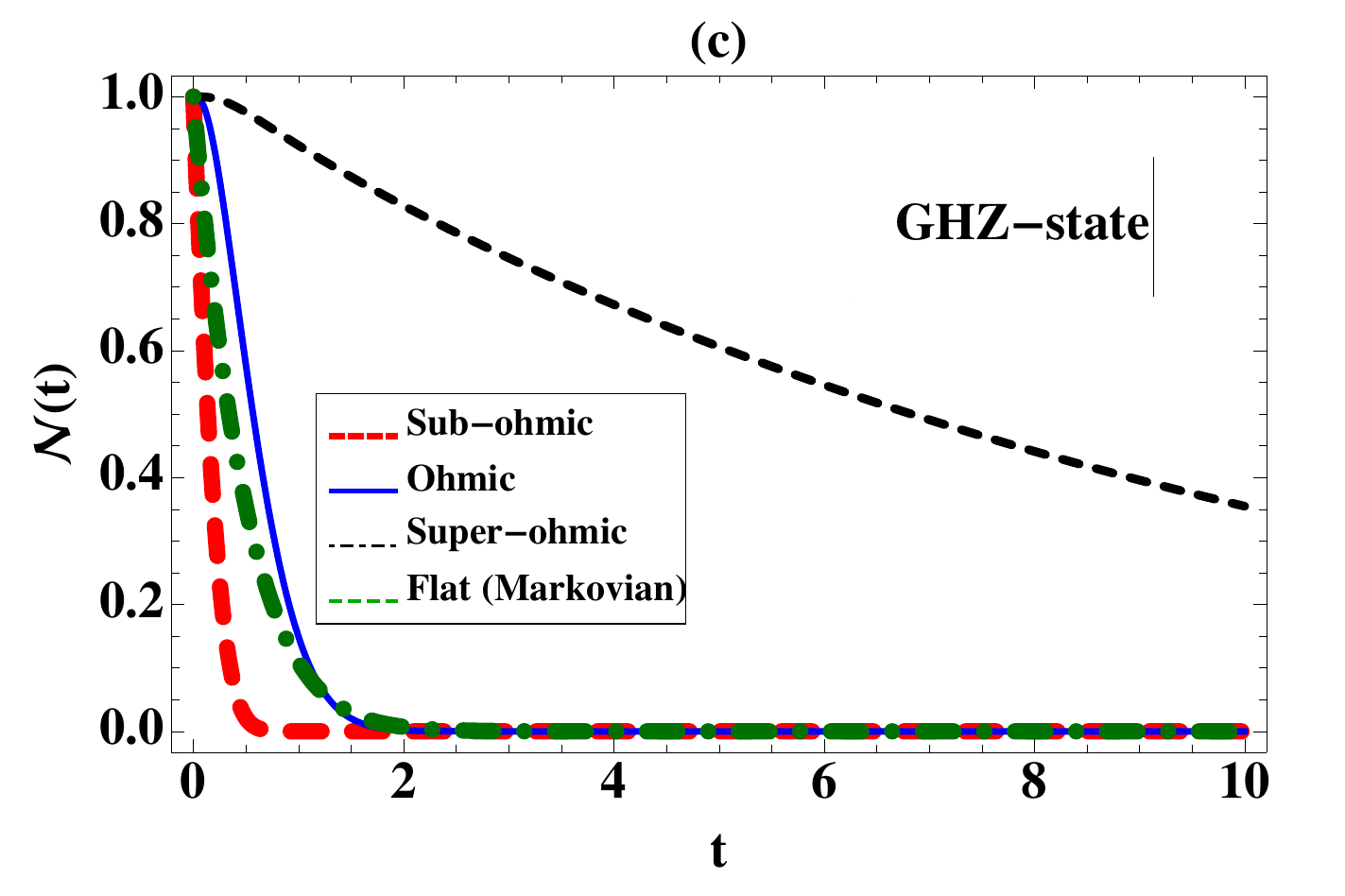}}
    \captionsetup{
  format=plain,
  margin=1em,                          
  justification=raggedright,
  singlelinecheck=false
}
  \caption{Plot of $\mathcal{N}(t)$ versus $t$ when an initial (a) W and (c) GHZ states are exposed to sub-ohmic ($s = 1/2$), ohmic ($s \rightarrow 1$) and super-ohmic ($s = 3$) non-Markovian baths. Note that at exactly $s = 1$ the time-dependent function $\kappa(t)$ blows up so to consider ohmic case we have taken the limit $s\rightarrow 1$. Other parameters used are: $\alpha_{sub} = 0.1, \omega_{cut} = 2$ for sub-ohmic, $\alpha_{ohm} = 0.6, \omega_{cut} = 10$ for ohmic and $\alpha_{sup} = 1, \omega_{cut} = 15$ for super-ohmic models and time axis is measured in $\alpha^{-1}_{sup}$ units. Figure inset shows time-dependent decay rate for super-ohmic bath.}\label{Fig7}
\end{figure*}

\subsection{Finite temperatures} The theoretical model described above can easily be extended to incorporate the consequences of finite temperatures. For simplicity here we'll present our results for the case of non-zero temperature Lorentzian  types of non-Markovian baths at low temperatures i.e. $\hbar\omega_{ci}>k_{B}T_{i}=1/\beta_{i}$. For the Lorentzian types of baths we find that under this temperature regime, time-dependent function (with assumption that $\alpha_{1}(t)=\alpha_{2}(t)=\alpha_{3}(t)\equiv\alpha(t)$ and $\beta_{1}(t)=\beta_{2}(t)=\beta_{3}(t)\equiv\beta(t)$ ) take the following form:

 \begin{subequations}
 \begin{align}
\alpha(t)=\nonumber \\
\frac{\alpha_{L}\Gamma^{2}\overline{N}}{2(\delta^{2}+\Gamma^{2})}\Bigg(e^{i\delta t}-1\Bigg)+\frac{\alpha_{L}\Gamma^{2}\overline{N}e^{-i\hbar\beta\Gamma}}{2\Gamma(\delta+i\Gamma)}\Bigg(1-e^{-\Gamma t}e^{-i\delta t}\Bigg),
\end{align}\\
\vspace{-8mm}
\begin{align}
\beta(t)=\alpha(t)+\frac{\alpha_{L}\Gamma}{2}\Bigg(\frac{1-e^{-\Gamma t}e^{-i\delta t}}{\Gamma+i\delta}\Bigg),
\end{align}
 \end{subequations}

Corresponding time-dependent decay rate $\kappa(t)$ which for the simple case of zero detunnings ($\delta=0$) turns out to be:
\begin{equation}
\kappa(t)=\alpha_{L}(1-e^{-\Gamma t})(\overline{N}cos(\hbar\beta\Gamma)+1)
\end{equation}
With these time-dependent functions we plot the negativity (for both initial W and GHZ states) in Fig.~8. In all plots we have taken $\overline{N}=0.1$. Compared to Fig.~3 (in which $\overline{N}=0$) we find that main effect of non-zero temperature is an enhanced decays of $\mathcal{N}(t)$ in all single, -double and -band-gap Lorentzian models. For example, in double Lorentzian case $\mathcal{N}(t)$ now dies out at around $\sim5\omega^{-1}_{c}$ while in $\overline{N}=0$ it died at $\sim6\omega^{-1}_{c}$. Rest of the features including the overall temporal profile, order of highest to lowest prolonging time periods of $\mathcal{N}(t)$ in different Lorentzian models and no-revival behavior remains the same as noticed in $\overline{N}=0$ case. Most interestingly, when $\overline{N}=0.1$ case for both Markovian (Fig.~2) and non-Markovian case (Fig.~8) is compared, one finds that even at non-zero temperatures, non-Markovian baths works better in tripartite entanglement storage (Markovian case $\mathcal{N}(t)$ dies at $1.25\kappa^{-1}$ while non-Markovian baths sustains $\mathcal{N}(t)$ at least till $2.5\omega^{-1}_{c}$ while $\kappa=\omega^{-1}_{c}$). Furthermore, we checked the same behavior extends down to $\delta \neq 0$ and lower $\alpha_{L}$ values cases as well (not shown here).
\begin{figure*}[t]
\centering
  \subfloat{%
    \includegraphics[width=5.4cm,height=4.4cm]{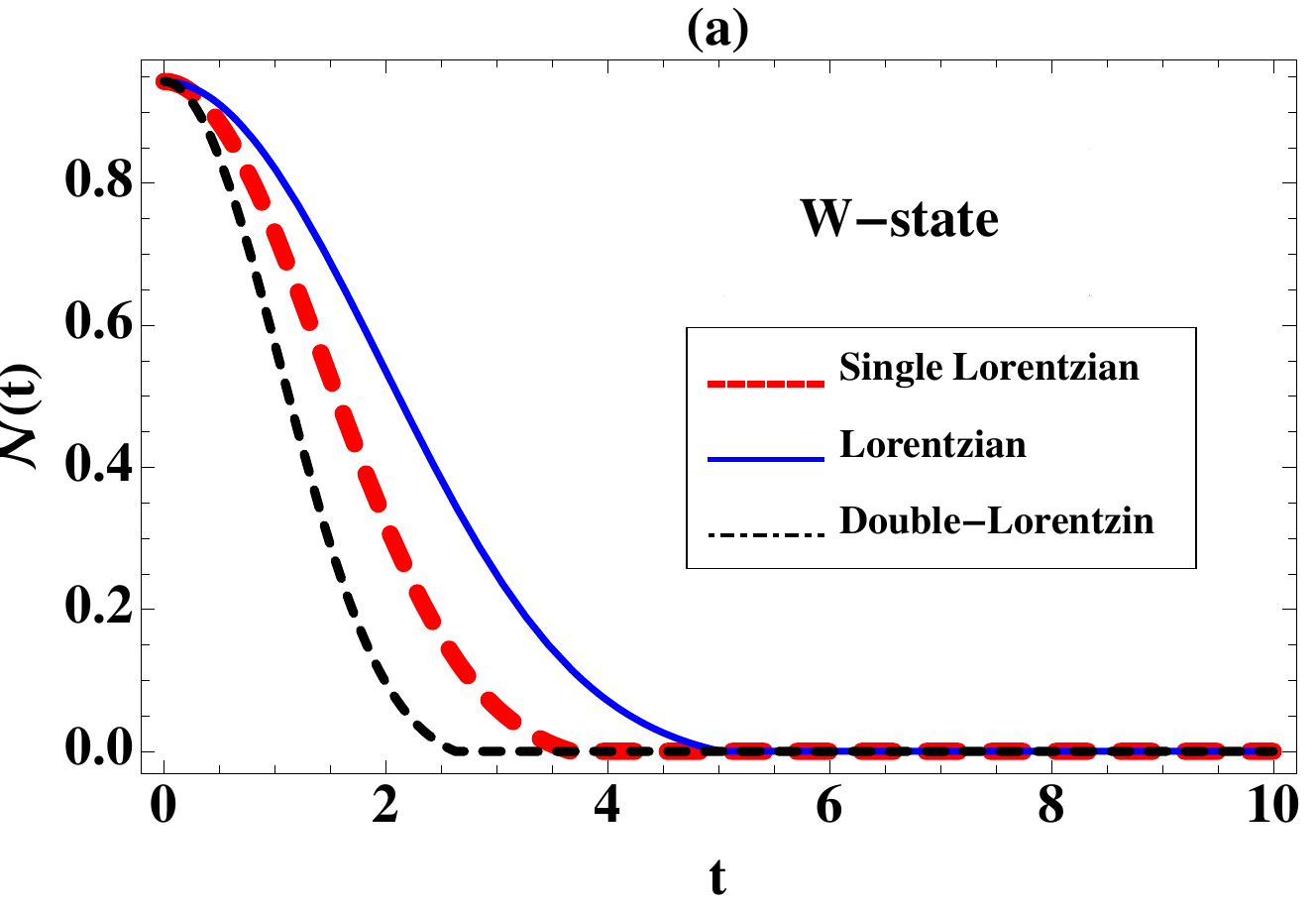}}
  \subfloat{%
    \includegraphics[width=5.4cm,height=4.4cm]{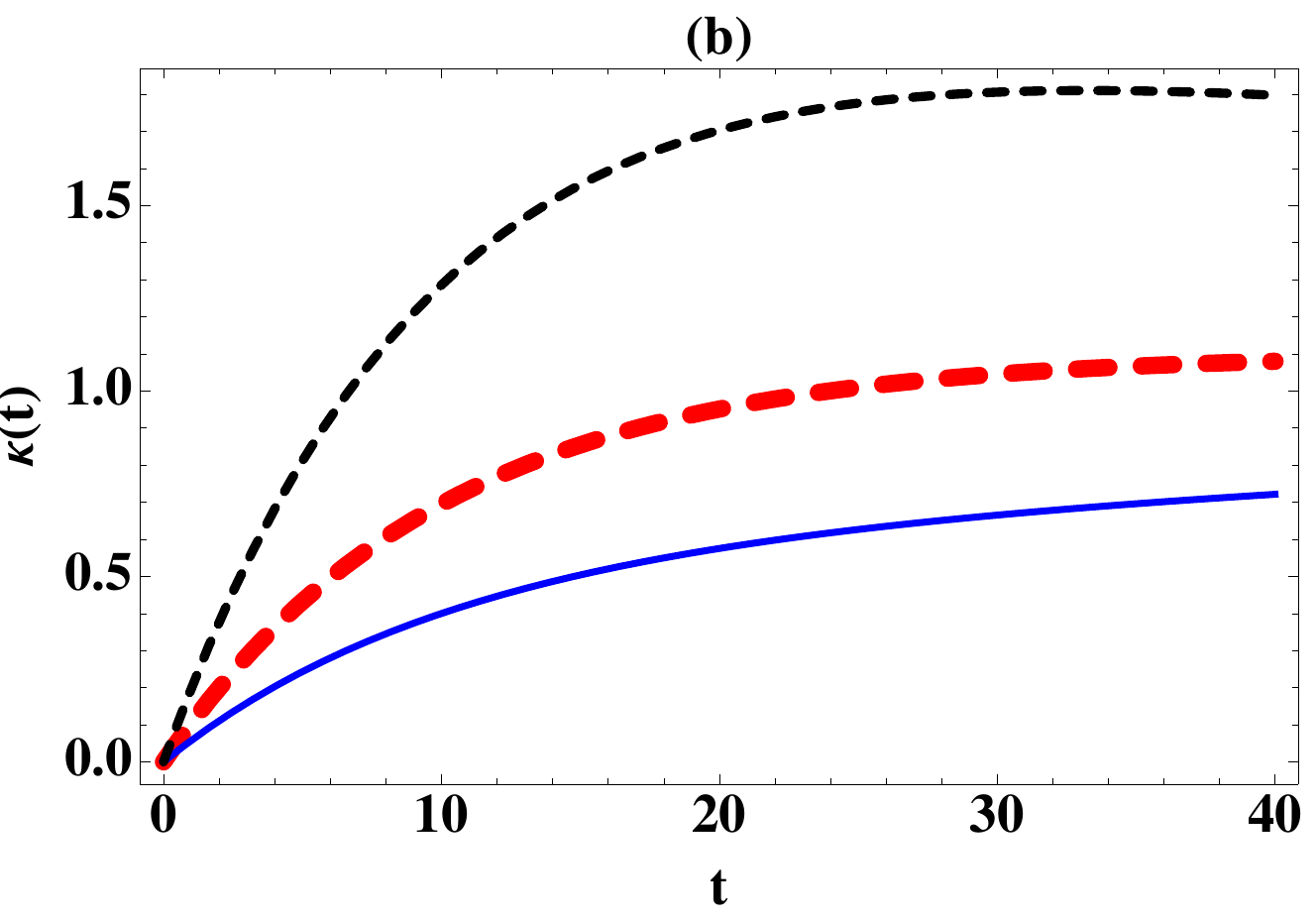}}
    \subfloat{%
    \includegraphics[width=5.4cm,height=4.4cm]{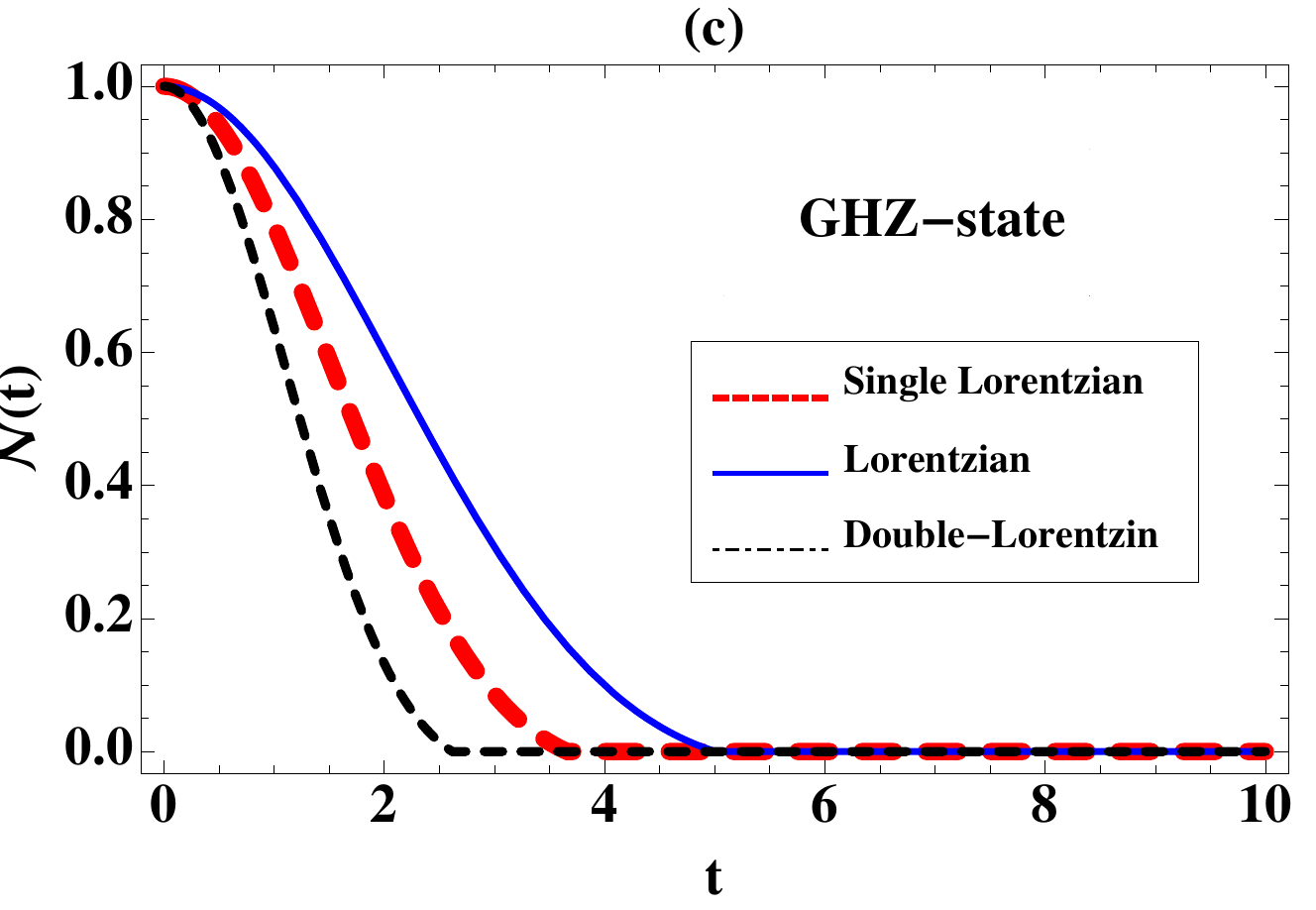}}
    \captionsetup{
  format=plain,
  margin=1em,                          
  justification=raggedright,
  singlelinecheck=false
}
  \caption{Finite temperature effect on non-Markovian tripartite entanglement. Low temperature regime with $\overline{N}=0.1$ and the rest of parameters same as in Fig.~3 with time measured in $\omega^{-1}_{c}$ units. As expected, inclusion of finite temperature in non-Markovian baths resulted in a relatively fast decay of $\mathcal{N}(t)$ in all cases. We notive that in finite temperatures case $\kappa(t)$ attains maximum value rather quickly than the corresponding zero temperature case which results in an earlier entanglement decay.}\label{Fig8}
\end{figure*}
\vspace{-4mm}
\section{Conclusions}
We showed by utilizing the technique of reservoir engineering through the use of non-Markovian baths (at zero and non-zero temperatures) with Lorentzian and ohmic type of spectral denisties, tripartite entanglement among optical cavities can be sustained for prolonged times. Either cavities start off in a maximally entangled W or GHZ state, we demonstrated that by probing the non-Markovian bath parameters (detunning, coupling strength and temperature) temporal profile of entanglement can be controlled and made more robust against effects of decoherence in comparison to the corresponding Markovian baths (with or without the possibility of photon hoping among cavities). In Lorentzian type of models we found far detunned double Lorentzian model with small coupling strengths, whereas in ohmic type of environments  super-ohmic baths with small cut-off frequencies are excellent candidates that can maintain the three-cavity entanglement for at least thrice longer times (almost $10\kappa^{-1}$) compared to all other Markovian and non-Markovian baths.\\
Along with the calculation of non-Markovian tripartite entanglement with different environmental models, the main novelty of this work is the application of NMQJA to understand the dynamical features of $\mathcal{N}(t)$ under non-zero detunning. We also substantiated that according to NMQJA, time-dependent decay rates are sufficient for the understanding of the entanglement collapse and revival. For the cases when time-dependent decay rate remains positive, an average value of $\kappa(t)$ can be used to describe the faster/slower decline of negativity as a function of time in different non-Markovian models. 
\section*{Acknowledgment}
I. M. M. would like to thank Jykri Piilo for useful comments and National Center for Physics, Islamabad, Pakistan for providing resources to complete this research.
\vspace{-4mm}
\appendix
\section{Equations of motions of density matrix elements} \label{App:AppendixA}
Note that the equations outlined below can describe the Markovian baths (with and without photon hoping) as well as the non-Markovian baths (with and without zero temperatures). By setting the $J_{i}(\omega)=\kappa_{i}/2\pi$ and $\overline{N}_{i}(\omega)=\overline{N}_{i}$ below equations describe the Markovian bath situation. While taking the time-dependent functions $\alpha_{i}(t) = 0$ and $\beta_{i}(t)=\int_{0}^{t}dt_{1}\int_{0}^{\infty}J_{i}(\omega)e^{i(\omega-\omega_{ci})(t-t_{1})}d\omega$ we enter into the non-Markovian bath regime with baths at zero temperatures. Photon hoping can be removed by considering $\xi_{12}=\xi_{23}=0$. Upper dash represents time derivative of matrix elements.  
\begin{gather}
\rho _{11}'(t)=2 \Re(\beta ) \left(\rho _{22}(t)+\rho _{33}(t)+\rho _{44}(t)\right)-6\Re(\alpha ) \rho _{11}(t)\\
\rho _{12}'(t)=\rho _{12}(t) \left(\alpha -\beta ^*-6\Re(\alpha )\right)+2 \Re(\beta ) \left(\rho _{35}(t)+\rho _{46}(t)\right)\\
\rho _{13}'(t)=i \xi _{12} \rho _{12}(t)+i \xi _{23} \rho _{14}(t)+\rho _{13}(t) \left(\alpha -\beta ^*-6\Re(\alpha )\right)\nonumber\\+2 \Re(\beta ) \left(\rho _{25}(t)+\rho _{47}(t)\right)\\
\rho _{14}'(t)=\rho _{14}(t) \left(\alpha - \beta ^*-6\Re(\alpha )\right)+2 \Re(\beta ) \left(\rho _{26}(t)+\rho _{37}(t)\right)\\
\rho _{15}'(t)=\rho _{15}(t) \left(2\alpha -2\beta ^*-6\Re(\alpha )\right)+2 \Re(\beta ) \rho _{48}(t)\\
\rho _{16}'(t)=i \xi _{23} \rho _{15}(t)+i \xi _{12} \rho _{17}(t)+\rho _{16}(t) \left(2\alpha -2\beta ^*-6\Re(\alpha )\right)\nonumber\\+2 \Re(\beta ) \rho _{38}(t)\\
\rho _{17}'(t)=\rho _{17}(t) \left(2\alpha -2\beta ^*-6\Re(\alpha )\right)+2 \Re(\beta ) \rho _{28}(t)\\
\rho _{18}'(t)=\rho _{18}(t) \left(-3\alpha ^*-3\beta ^*\right)\\
\rho _{21}'(t)=\rho _{21}(t) (-\alpha -\beta -4\Re(\alpha ))+2 \Re(\beta ) \left(\rho _{53}(t)+\rho _{64}(t)\right)\\
\rho _{22}'(t)=i \xi _{12} \left(\rho _{23}(t)-\rho _{32}(t)\right)+\rho _{22}(t) (-4\Re(\alpha )-2\Re(\beta ))\nonumber\\+2 \Re(\alpha ) \rho _{11}(t)+2 \Re(\beta ) \rho _{55}(t)
+2 \Re(\beta ) \rho _{66}(t)\\
\rho _{23}'(t)=i \xi _{23} \rho _{24}(t)\nonumber\\-i \xi _{12} \rho _{22}(t)+\rho _{23}(t) (-4\Re(\alpha )-2\Re(\beta ))+2\Re(\beta ) \rho _{67}(t)\\
\rho _{24}'(t)=\rho _{24}(t) (-4\Re(\alpha )-2\Re(\beta ))+2\Re(\beta ) \rho _{57}(t)\\
\rho _{25}'(t)=\rho _{25}(t) (\alpha +\beta -4\Re(\alpha )-4\Re(\beta ))+\nonumber\\2 \Re(\alpha ) \rho _{13}(t)+2 \Re(\beta ) \rho _{68}(t)\\
\rho _{26}'(t)=i \xi _{23} \rho _{25}(t)+i \xi _{12} \rho _{27}(t)-i \xi _{12} \rho _{36}(t)+\rho _{26}(t) (\alpha +\beta\nonumber\\ -4\Re(\alpha )-4\Re(\beta ))\\
+2 \Re(\alpha ) \rho _{14}(t)+2 \Re(\beta ) \rho _{58}(t)\nonumber\\
\rho _{27}'(t)=\rho _{27}(t) (\alpha +\beta -4\Re(\alpha )-4\Re(\beta ))\\
\rho _{28}'(t)=\rho _{28}(t) \left(2\beta -2\alpha ^*-6\Re(\beta )\right)+2 \Re(\alpha ) \rho _{17}(t)\\
\rho _{31}'(t)=\rho _{31}(t) (-\alpha -\beta -4\Re(\alpha ))+2 \Re(\beta ) \left(\rho _{52}(t)+\rho _{74}(t)\right)\\
\rho _{32}'(t)=\rho _{32}(t) (-4\Re(\alpha )-2\Re(\beta ))+2\Re(\beta ) \rho _{76}(t)\\
\rho _{33}'(t)=-i \xi _{12} \rho _{23}(t)+i \xi _{12} \rho _{32}(t)+i \xi _{23} \rho _{34}(t)-i \xi
  _{23} \rho _{43}(t)\nonumber\\+\rho _{33}(t) (-4\Re(\alpha )-2\Re(\beta ))
  +2 \Re(\alpha ) \rho _{11}(t)\nonumber\\+2 \Re(\beta )\rho _{55}(t)+2 \Re(\beta ) \rho _{77}(t)
      \end{gather}
\begin{gather}
\rho _{34}'(t)=\rho _{34}(t) (-4\Re(\alpha )-2\Re(\beta ))+2\Re(\beta ) \rho _{56}(t)\\
\rho _{35}'(t)=\rho _{35}(t) (\alpha +\beta -4\Re(\alpha )-4\Re(\beta ))+2 \Re(\alpha ) \rho _{12}(t)\nonumber\\+2 \Re(\beta ) \rho _{78}(t)\\
\rho _{36}'(t)=-i \xi _{12} \rho _{26}(t)+i \xi _{23} \rho _{35}(t)+i \xi _{12} \rho _{37}(t)+i \xi
   _{23} \rho _{46}(t)+\nonumber\\\rho _{36}(t) (\alpha +\beta -4\Re(\alpha )-4\Re(\beta ))\\
\rho _{37}'(t)=\rho _{37}(t) (\alpha +\beta -4\Re(\alpha )-4\Re(\beta ))+2 \Re(\alpha ) \rho _{14}(t)\nonumber\\+2 \Re(\beta ) \rho _{58}(t)\\
\rho _{38}'(t)=\rho _{38}(t) \left(2\beta -2\alpha ^*-6\Re(\beta )\right)+2 \Re(\alpha ) \rho _{16}(t)\\
\rho _{41}'(t)=\rho _{41}(t) (-\alpha -\beta -4\Re(\alpha ))+2 \Re(\beta ) \left(\rho _{62}(t)+\rho _{73}(t)\right)\\
\rho _{42}'(t)=\rho _{42}(t) (-4\Re(\alpha )-2\Re(\beta ))+2\Re(\beta ) \rho _{75}(t)\\
\rho _{43}'(t)=-i \xi _{23} \rho _{33}(t)+i \xi _{12} \rho _{42}(t)+i \xi _{23} \rho _{44}(t)+\nonumber\\\rho _{43}(t) (-4\Re(\alpha )-2\Re(\beta ))+2\Re(\beta ) \rho _{65}(t)\\
\rho _{44}'(t)=\rho _{44}(t) (-4\Re(\alpha )-2\Re(\beta ))+2 \Re(\alpha ) \rho _{11}(t)+\nonumber\\2 \Re(\beta ) \rho _{66}(t)+2 \Re(\beta ) \rho _{77}(t)\\
\rho _{45}'(t)=\rho _{45}(t) (\alpha +\beta -4\Re(\alpha )-4\Re(\beta ))\\
\rho _{46}'(t)=-i \xi _{23} \rho _{36}(t)+i \xi _{23} \rho _{45}(t)+i \xi _{12} \rho _{47}(t)\nonumber\\+\rho _{46}(t) (\alpha +\beta -4\Re(\alpha )-4\Re(\beta ))\\
\hspace{-10mm}+2 \Re(\alpha ) \rho _{12}(t)+2 \Re(\beta ) \rho _{78}(t)\nonumber\\
\rho _{47}'(t)=\rho _{47}(t) (\alpha +\beta -4\Re(\alpha )-4\Re(\beta ))\nonumber\\+2 \Re(\alpha ) \rho _{13}(t)+2 \Re(\beta ) \rho _{68}(t)\\
\rho _{48}'(t)=\rho _{48}(t) \left(2\beta -2\alpha ^*-6\Re(\beta )\right)+2 \Re(\alpha ) \rho _{15}(t)\\
\rho _{51}'(t)=\rho _{51}(t) (-2\alpha -2\beta -2\Re(\alpha ))+2 \Re(\beta ) \rho _{84}(t)\\
\rho _{52}'(t)=\rho _{52}(t) (-\alpha -\beta -2\Re(\alpha )-2\Re(\beta ))+\nonumber\\2 \Re(\alpha ) \rho _{31}(t)+2 \Re(\beta ) \rho _{86}(t)\\
\rho _{53}'(t)=-i \xi _{23} \rho _{36}(t)+i \xi _{12} \rho _{52}(t)+i \xi _{23} \rho _{54}(t)\nonumber\\+\rho _{53}(t) (-\alpha -\beta -2\Re(\alpha )-2\Re(\beta ))\\
+2 \Re(\alpha ) \rho _{21}(t)+2 \Re(\beta ) \rho _{87}(t)\nonumber\\
\rho _{54}'(t)=\rho _{54}(t) (-\alpha -\beta -2\Re(\alpha )-2\Re(\beta ))\\
\rho _{55}'(t)=\rho _{55}(t) (-2\Re(\alpha )-4\Re(\beta ))+2 \Re(\alpha ) \rho _{22}(t)+\nonumber\\2 \Re(\alpha ) \rho _{33}(t)+2 \Re(\beta ) \rho _{88}(t)\\
\rho _{56}'(t)=i \xi _{23} \rho _{55}(t)+i \xi _{12} \rho _{57}(t)-i \xi _{23} \rho _{66}(t)+\nonumber\\\rho _{56}(t) (-2\Re(\alpha )-4\Re(\beta ))+2 \Re(\alpha ) \rho _{34}(t)\\
\rho _{57}'(t)=\rho _{57}(t) (-2\Re(\alpha )-4\Re(\beta ))+2 \Re(\alpha ) \rho _{24}(t)\\
\rho _{58}'(t)=\rho _{58}(t) \left(\beta -\alpha ^*-6\Re(\beta )\right)+2 \Re(\alpha ) \left(\rho _{26}(t)+\rho _{37}(t)\right)\\
\rho _{61}'(t)=\rho _{61}(t) (-2\alpha -2\beta -2\Re(\alpha ))+2 \Re(\beta ) \rho _{83}(t)\\
\rho _{62}'(t)=\rho _{62}(t) (-\alpha -\beta -2\Re(\alpha )-2\Re(\beta ))+\nonumber\\2 \Re(\alpha ) \rho _{41}(t)+2 \Re(\beta ) \rho _{85}(t)\\
\rho _{63}'(t)=-i \xi _{23} \rho _{53}(t)+i \xi _{12} \rho _{62}(t)+i \xi _{23} \rho _{64}(t)\nonumber\\-i \xi
   _{12} \rho _{73}(t)+\rho _{63}(t) (-\alpha -\beta -2\Re(\alpha )-2\Re(\beta ))
    \end{gather}
\begin{gather}
\rho _{64}'(t)=\rho _{64}(t) (-\alpha -\beta -2\Re(\alpha )-2\Re(\beta ))+\nonumber\\2 \Re(\alpha ) \rho _{21}(t)+2 \Re(\beta ) \rho _{87}(t)\\
\rho _{65}'(t)=\rho _{65}(t) (-2\Re(\alpha )-4\Re(\beta ))+2 \Re(\alpha ) \rho _{43}(t)\\
\rho _{66}'(t)=-i \xi _{23} \rho _{56}(t)+i \xi _{23} \rho _{65}(t)+i \xi _{12} \rho _{67}(t)-\nonumber\\i \xi
   _{12} \rho _{76}(t)+\rho _{66}(t) (-2\Re(\alpha )-4\Re(\beta ))\\
   +2 \Re(\alpha ) \rho _{22}(t)+2 \Re(\alpha ) \rho _{44}(t)+2 \Re(\beta ) \rho _{88}(t)\nonumber\\
\rho _{67}'(t)=\rho _{67}(t) (-2\Re(\alpha )-4\Re(\beta ))+2 \Re(\alpha ) \rho _{23}(t)\\
\rho _{68}'(t)=\rho _{68}(t) \left(\beta -\alpha ^*-6\Re(\beta )\right)+2 \Re(\alpha ) \left(\rho _{25}(t)+\rho _{47}(t)\right)\\
\rho _{71}'(t)=\rho _{71}(t) (-2\alpha -2\beta -2\Re(\alpha ))+2 \Re(\beta ) \rho _{82}(t)\\
\rho _{72}'(t)=\rho _{72}(t) (-\alpha -\beta -2\Re(\alpha )-2\Re(\beta ))\\
\rho _{73}'(t)=-i \xi _{12} \rho _{63}(t)+i \xi _{12} \rho _{72}(t)+i \xi _{23} \rho _{74}(t)+\nonumber\\\rho _{73}(t) (-\alpha -\beta -2\Re(\alpha )-2\Re(\beta ))\\
+2 \Re(\alpha ) \rho _{41}(t)+2 \Re(\beta ) \rho _{85}(t)\nonumber \\  
\rho _{74}'(t)=\rho _{74}(t) (-\alpha -\beta -2\Re(\alpha )-2\Re(\beta ))+\nonumber\\2 \Re(\alpha ) \rho _{31}(t)+2 \Re(\beta ) \rho _{86}(t)\\
\rho _{75}'(t)=\rho _{75}(t) (-2\Re(\alpha )-4\Re(\beta ))+2 \Re(\alpha ) \rho _{42}(t)\\
\rho _{76}'(t)=-i \xi _{12} \rho _{66}(t)+i \xi _{12} \rho _{75}(t)+i \xi _{12} \rho _{77}(t)+\nonumber\\\rho _{76}(t) (-2\Re(\alpha )-4\Re(\beta ))+2 \Re(\alpha ) \rho _{32}(t)\\
\rho _{77}'(t)=\rho _{77}(t) (-2\Re(\alpha )-4\Re(\beta ))+2 \Re(\alpha ) \rho _{33}(t)+\nonumber\\2 \Re(\alpha ) \rho _{44}(t)+2 \Re(\beta ) \rho _{88}(t)\\
\rho _{78}'(t)=\rho _{78}(t) \left(\beta -\alpha ^*-6\Re(\beta )\right)+2 \Re(\alpha ) \left(\rho _{35}(t)+\rho _{46}(t)\right)\\
\rho _{81}'(t)=(-3\alpha -3\beta ) \rho _{81}(t)\\
\rho _{82}'(t)=\rho _{82}(t) (-2\alpha -2\beta -2\Re(\beta ))+2 \Re(\alpha ) \rho _{71}(t)\\
\rho _{83}'(t)=i \xi _{12} \rho _{82}(t)+i \xi _{23} \rho _{84}(t)+\rho _{83}(t) (-2\alpha -2\beta -2\Re(\beta ))\nonumber\\+2 \Re(\alpha ) \rho _{61}(t)\\
\rho _{84}'(t)=\rho _{84}(t) (-2\alpha -2\beta -2\Re(\beta ))+2 \Re(\alpha ) \rho _{51}(t)\\
\rho _{85}'(t)=\rho _{85}(t) (-\alpha -\beta -4\Re(\beta ))+2 \Re(\alpha ) \left(\rho _{62}(t)+\rho _{73}(t)\right)\\
\rho _{86}'(t)=i \xi _{23} \rho _{85}(t)+i \xi _{12} \rho _{87}(t)+\rho _{86}(t) (-\alpha -\beta -4\Re(\beta ))\nonumber\\+2 \Re(\alpha ) \left(\rho _{52}(t)+\rho _{74}(t)\right)\\
\rho _{87}'(t)=\rho _{87}(t) (-\alpha -\beta -4 \Re(\beta ))+\nonumber\\2 \Re(\alpha ) \left(\rho _{53}(t)+\rho _{64}(t)\right)\\
\rho _{88}'(t)=2 \Re(\alpha ) \left(\rho _{55}(t)+\rho _{66}(t)+\rho _{77}(t)\right)-6\Re(\beta ) \rho _{88}(t)
\end{gather}

\section{Review of non-Markovian Quantum Jump Approach} \label{App:AppendixB}
Closely following the reference \cite{piilo2008non}, in this section we briefly review the NMQJA/quantum Monte Carlo wave function method introduced by Piilo et. al. In standard Markovian QJA we think of actual/fictitious photo detectors surrounding the system of interest such that whenever a photon is lost by the system any one of the detectors register it by making a click. According to the QJA then in any given small time interval we have one of two situations:

{\bf(I) Jump Case}: A quantum jump takes place, and we apply the output (annihilation) operator $\hat{J}_{j}$ (\cite{gardiner2004quantum}) associated with the jth jump channel/detector (which in this context is also termed the ``jump operator'') to the state of the system.

{\bf(II) No Jump Case}: No detector clicks, the system evolves according to the following non-Unitary Schr\"odinger equation:
\begin{equation}\label{NUSE}
i\hbar\frac{d}{dt}\ket{\tilde{\Psi}(t)}=\hat{H}_{NH}\ket{\tilde{\Psi}(t)},
\end{equation}
where the ``Hamiltonian'' appearing in this equation is a non-Hermitian operator, which is the sum of two parts. The first part is Hermitian and is given by the system's Hamiltonian, the second part is anti-Hermitian and  is constructed from the jump operators. In total we have
\begin{equation}\label{NHH}
\hat{H}_{NH}=\hat{H}_{sys}-\frac{i\hbar}{2}\sum_{j}\gamma_{j}\hat{J}^{\dagger}_{j}\hat{J}_{j}
\end{equation}
$\gamma_{j}$ is the rate at which excitation decays through the jth channel. Note that for the sake of clarity here instead of using $\kappa$ we are using $\gamma$ as decay functions. Based on the cycle of these oscillations jump part of the non-Hermitian Hamiltonian is then divided into negative and positive parts as:
\begin{equation}\label{NHHPN}
\hat{H}_{NH}=\hat{H}_{sys}-\frac{i\hbar}{2}\sum_{j+}\gamma_{j+}(t)\hat{J}^{\dagger}_{j+}\hat{J}_{j+}-\frac{i\hbar}{2}\sum_{j-}\gamma_{j-}(t)\hat{J}^{\dagger}_{j-}\hat{J}_{j-}
\end{equation}
Note that corresponding to positive and negative decay rates $\gamma_{j\pm}(t)$, positive and negative jump operators $\hat{J}_{j\pm}$ are introduced. Interpretation of these negative and positive jump channels is as follows:\\
During the positive channel/cycle of $\gamma_{j}(t)$ ($\gamma_{j}(t) > 0$), system behaves as the source of the excitation and the environment becomes the target of the information flow. Hence after the positive jump system loses information thus producing the state reduction in an abrupt way as:
\begin{equation}
\ket{\Psi_{\alpha}}\longrightarrow \ket{\Psi_{\alpha'}}=\frac{\hat{J}_{j+}\ket{\Psi_{\alpha}}}{||\hat{J}_{j+}\ket{\Psi_{\alpha}}||}
\end{equation}
subscript $\alpha$ is introduced to mention that at time t there are $N_{\alpha}$ number of ensemble members in the source state $\ket{\Psi_{\alpha}(t)}$. $\ket{\Psi_{\alpha'}(t)}$ is the target state in this scenario with $N_{\alpha'}$ members. Occurrence of jump is a random process which in the positive decay cycle happens in a small time interval $[t,t+\delta t]$ with the probability:
\begin{equation}
P_{\alpha}^{j+}(t) = \gamma_{j+}(t)\langle \Psi_{\alpha}(t)\vert \hat{J}_{j+}^{\dagger}\hat{J}_{j+}\vert \Psi_{\alpha}(t)\rangle \delta t
\end{equation}
 Note that the Markovian QJA is a special case of this positive decay cycle situation with positive time-independent decay rates.\\
 During the negative cycle/channel ($\gamma_{j-}(t)<0$) jump process reverses it's direction i.e. memory-full environment becomes the source and  system acts as the target of the information
\begin{equation}
\ket{\Psi_{\alpha'}}\longleftarrow \ket{\Psi_{\alpha}}=\frac{\hat{J}_{j-}\ket{\Psi_{\alpha'}}}{||\hat{J}_{j-}\ket{\Psi_{\alpha'}}||}
\end{equation}
It is worth mentioning here that click description of quantum jumps is (strictly speaking) valid for positive jumps only. For negative jumps, the click picture is still an under debate issue (for further details see for instance \cite{wiseman2008pure,diosi2008non}). During negative cycle, probability of negative/non-Markovian jump turns out to be:
\begin{equation}
P_{\alpha\rightarrow\alpha'}^{j-}(t) = \frac{N_{\alpha'}(t)}{N_{\alpha}(t)}|\gamma_{j-}(t)|\langle \Psi_{\alpha'}(t)\vert \hat{J}_{j-}^{\dagger}\hat{J}_{j-}\vert \Psi_{\alpha'}(t)\rangle \delta t
\end{equation}
Note that the negative jumps are actually responsible for producing memory effects (non-Markovianity) in the system dynamics. In above equation negative jump probability is weighted by the ratio of number of ensemble members in the target/system state to the number of members in the source/environment state. This implies that negative jump probability is going to be zero if there are no members in the target state. Moreover we have taken the absolute value of negative decay rate to ensure that the probability remains non-negative. Sign of the negative decay (as already mentioned) describes the reverse flow of information and therefore non-Markovian quantum jump approach doesn't suffer the issue of negative probabilities \cite{breuer2002theory}. Finally note that one can take the ensemble average over all different possible trajectories (no, positive and negative jump situations) to construct the full density matrix which will be equivalent to the density matrix obtained by directly solving the full non-Markovian Master equations \cite{piilo2008non}.

\bibliography{Article}
\end{document}